\documentclass[
twocolumn,secnumarabic
,amssymb, amsmath,nobibnotes, aps, prl]{revtex4}
\usepackage[dvips]{graphicx}
\usepackage{tabularx}
\usepackage{bm}%
\expandafter\ifx\csname package@font\endcsname\relax\else
 \expandafter\expandafter
 \expandafter\usepackage
 \expandafter\expandafter
 \expandafter{\csname package@font\endcsname}%
\fi

\begin{document}

\title{The ${\rm AdS}^2_{\theta}/{\rm CFT}_1$ Correspondence and Noncommutative Geometry II:\\
Noncommutative Quantum Black Holes}

\author{Badis Ydri}
\affiliation{Department of Physics, Badji-Mokhtar Annaba University,\\
 Annaba, Algeria.}

\begin{abstract}
In this article we present the construction of noncommutative ${\rm AdS}^2_{\theta}$ black hole and its four-dimensional Yang-Mills IKKT-type matrix model which includes two competing Myers term one responsible for the condensation of pure ${\rm AdS}^2_{\theta}$ and the other one responsible for the condensation of the dilaton field. It is argued that the phase diagram of this matrix model features three phases: 1) A
gravitational phase (${\rm AdS}^2_{\theta}$ black hole), 2) A geometric phase (${\rm AdS}^2_{\theta}$ background) and 3) A Yang-Mills phase. The Hawking process is therefore seen as an exotic line of discontinuous transitions between the gravitational and geometrical phases. Alternatively, a noncommutative non-linear sigma model describing the transition of the dilaton field between the gravitational and geometrical phases is also constructed.

\end{abstract}

\maketitle
\tableofcontents
\section{Introduction}

In this article a synthesis of the principles of noncommutative geometry and their matrix models together with the principles of the ${\rm AdS}^{d+1}/{\rm CFT}_d$ correspondence for $d=1$ is presented.

The original motivation behind this work comes from fuzzy physics \cite{Ydri:2001pv}  as well as from non-perturbative lattice-like approaches to superstring theory which are matrix-based \cite{Hanada:2016jok}.

\subsection{A ${\rm QM}/{\rm NCG}$ Correspondence}
In the first part of our study \cite{ydri} we have put forward a novel proposal for the  ${\rm AdS}^2/{\rm CFT}_1$ correspondence in which we turn the usual understanding on its head. It is customary to assume that the bulk theory is given by some gravity theory about an AdS background and then seek a quantum field theory residing on the boundary which is dual to the gravity theory in the bulk. As it happens the case of ${\rm AdS}^2$ is very different from higher dimensional anti-de Sitter spacetimes as the boundary theory is essentially unknown. See for example, \cite{Strominger:1998yg,Spradlin:1999bn} and \cite{Cadoni:1998sg,Cadoni:1999ja}.

This difficulty is usually traced to the fact that ${\rm AdS}^2$ has two disconnected one-dimensional boundaries as opposed to higher dimensional AdS which are characterized by a single connected boundary. But in here we trace instead this difficulty to the fact that ${\rm CFT}_1$ as given by conformal quantum mechanics is really  only quasi-conformal (in a sense to be {\it specified}) and as a consequence the theory in the bulk is only required to be quasi-AdS (in a sense to be {\rm proposed}). See for example,  \cite{deAlfaro:1976vlx,Chamon:2011xk} but also \cite{Okazaki:2015lpa,Okazaki:2017lpn,Gupta:2019cmo,Gupta:2017lwk,Gupta:2015uga,Gupta:2013ata}.

Thus, in this proposal we are turning our understanding of the ${\rm AdS}^{d+1}/{\rm CFT}_d$ on its head since we are starting from the boundary and then we are moving towards the bulk and not the other way around. We are thus insisting that on the boundary the ${\rm CFT}_1$ is  really given by dAFF conformal quantum mechanics. However, this conformal quantum mechanics is only "quasi-conformal" in the sense that there is neither an $SO(1,2)-$invariant quantum vacuum state nor there are primary operators in the strict sense yet the bulk correlators are correctly reproduced by appropriately defined boundary quantum fields.

The fact that we insist that the boundary theory is given by this quasi-conformal "conformal quantum mechanics" leads us to the conclusion that the bulk theory is not necessarily given by commutative   ${\rm AdS}^2$ which should strictly corresponds to conformal invariance. In fact,  we will argue that noncommutative  ${\rm AdS}^2_{\theta}$ is a much better candidate as it shares the same group structure as this quasi-conformal "conformal quantum mechanics" while it is quasi-AdS in the sense that it behaves more and more as commutative   ${\rm AdS}^2$ as we approach the boundary. Indeed, both commutative  ${\rm AdS}^2$ and noncommutative  ${\rm AdS}^2_{\theta}$ are characterized by the same boundary.

It is further observed that the Lorentz group $SO(1,2)$ is the fundamental unifying structure of the ${\rm AdS}^2$ spacetime, the noncommutative ${\rm AdS}^2_{\theta}$ space and of the boundary quantum theory. In particular, the algebra of quasi-primary operators at the boundary is seen to be a subalgebra of the operator algebra of  noncommutative ${\rm AdS}^2_{\theta}$. This leads us to the conclusion/conjecture that the theory in the bulk must be given by noncommutative geometry \cite{Connes:1996gi} and not by classical gravity, i.e. it is given by noncommutative ${\rm AdS}^2_{\theta}$ and not by commutative ${\rm AdS}^2$. Thus, the "quasi-conformal" dAFF conformal quantum mechanics on the boundary is actually dual to the "quasi-AdS"  noncommutative  ${\rm AdS}^2_{\theta}$ in the bulk.

\subsection{Noncommutative Quantum Black Holes}
In this article we present the second part of our study in which a synthesis of the principles of noncommutative geometry and the ${\rm AdS}^{2}/{\rm CFT}_1$ correspondence is outlined. In this part the main focus will be on constructing noncommutative ${\rm AdS}^2_{\theta}$ black holes and their matrix models.

First, we note that Connes' noncommutative geometry is understood here as "first quantization" of geometry whereas the corresponding matrix models (typically Yang-Mills matrix models of the IKKT-type \cite{Ishibashi:1996xs}) provide "quantum gravity" or "second quantization" of the corresponding geometry. 

We begin with a brief presentation of the differential geometry of  commutative ${\rm AdS}^2$ and its quantization into the noncommutative geometry of noncommutative ${\rm AdS}^2_{\theta}$. Then we present an overview of the $SO(1,2)-$coherent state which allows us to intuitively visualize noncommutative ${\rm AdS}^2_{\theta}$ and provides the natural starting point for the construction of the Weyl map and Moyal-Weyl star product and hence allows us a concrete handle on the commutative limit.

After presenting the noncommutative geometry of noncommutative ${\rm AdS}^2_{\theta}$ we present its IKKT-type Yang-Mills matrix models (which includes a cubic Myers term \cite{Myers:1999ps}). Then we briefly discuss the quantum stability of this noncommutative/matrix background and the UV-IR mixing of the gauge degrees of freedom. The detailed discussion of the phase structure is relegated to the forthcoming third part of our study. 

The commutative ${\rm AdS}^2$ black holes are then constructed in the two-dimensional Jackiw-Teitelboim dilaton gravity and the corresponding Hawking process \cite{Hawking:1974sw} is formulated. The scalar field  (the dilaton) plays a fundamental role here in distinguishing between pure ${\rm AdS}^2$ spacetime (${\rm AdS}_0$) and ${\rm AdS}^2$ black hole spacetime (${\rm AdS}_+$). Indeed, these two solutions are locally equivalent and they are only globally/topologically differentiated by the value of the dilaton field \cite{Cadoni:1993rn}. The Hawking process is then formulated between ${\rm AdS}_0$ (ground state) and ${\rm AdS}_+$ (excited state) as essentially an Unruh effect \cite{Unruh:1976db}.

These  ${\rm AdS}^2$ black hole configurations are then quantized (similarly to the quantization of pure ${\rm AdS}^2$ spacetime) to obtain noncommutative black hole configurations around noncommutative ${\rm AdS}^2_{\theta}$. Then we derive a four-dimensional Yang-Mills matrix model with mass deformations which "second quantize" these noncommutative ${\rm AdS}^2_{\theta}$ black hole configurations. This IKKT-type Yang-Mills matrix model includes two competing Myers term one responsible for the condensation of pure ${\rm AdS}^2_{\theta}$ and the other one responsible for the condensation of the dilaton field. It is argued that the phase diagram of this matrix model features three phases: 1) A gravitational phase (${\rm AdS}^2_{\theta}$ black hole), 2) A geometric phase (${\rm AdS}^2_{\theta}$ background) and 3) A Yang-Mills phase (no discernible gravitational/geometrical content in the sense described here).

The Hawking process within the Yang-Mills matrix model is therefore given by an exotic line of discontinuous transitions with a jump in the entropy, characteristic of a 1st order transition, yet with divergent critical fluctuations and a divergent specific heat, characteristic of a 2nd order transition, between the gravitational and geometrical phases.

Another proposal for noncommutative dilaton gravity is also put forward. In this alternative scenario dilaton gravity presents itself as a non-linear sigma model. The vacuum expectation value of the dilaton field should vanish in the pure ${\rm AdS}^2$ background while it will take a non-trivial value in the ${\rm AdS}^2$ black hole background. In the context of the noncommutative non-linear sigma model this transition is exhibited as a third order transition from a disordered one-cut solution (pure ${\rm AdS}^2_{\theta}$) to a uniform-ordered two-cut solution (${\rm AdS}^2_{\theta}$ black hole).

In this article we will also construct the star product in noncommutative black hole configurations using Darboux theorem. This will complete and complement our semi-classical treatment of noncommutative ${\rm AdS}^2_{\theta}$ and  noncommutative ${\rm AdS}^2_{\theta}$ black hole and their commutative limits.

We also include two appendices. The first appendix contains a much more detailed discussion of the Hawking process in ${\rm AdS}^2$ spacetime. The second appendix contains a brief discussion of the relation between gauge and coordinate transformations in the context of deformation quantization.

\subsection{Phase Structure of the Noncommutative ${\rm AdS}^2_{\theta}\times \mathbb{S}^2_N$}
In the forthcoming third part of our study, in which a synthesis of the principles of noncommutative geometry and the ${\rm AdS}^{2}/{\rm CFT}_1$ correspondence is put forward, we will derive explicitly the phase structure of the IKKT-type Yang-Mills matrix models describing the quantum gravitational fluctuations around the fuzzy sphere $\mathbb{S}^2_N$, the noncommutative pseudo-sphere $\mathbb{H}^2_{\theta}$, the noncommutative anti-de Sitter spacetime ${\rm AdS}^2_{\theta}$ and the noncommutative near-horizon geometry ${\rm AdS}^2_{\theta}\times\mathbb{S}^2_N$.

\section{The commutative ${\rm AdS}^2$}
The bulk geometry is defined by the metric of ${\rm AdS}^2$ given in the cylindrical coordinates $(\sigma,\tau)$ by 

 \begin{eqnarray}
   ds^2&=&\frac{R^2}{\cos^2 \sigma}(-d\tau^2+d\sigma^2).\label{met1}
 \end{eqnarray} 
The global coordinates of ${\rm AdS}^2$, which is a spacetime embedded in Minkowski spacetime  $\mathbb{R}^{2,1}$ with metric $\eta=(-1,-1,+1)$, consists of the two coordinates $\tau$ and $\sigma$. These coordinates are given explicitly by

\begin{eqnarray}
&&X_{1}=R\frac{\cos \tau}{\cos \sigma}=-\frac{z}{2}(1+\frac{R^2-t^2}{z^2})\nonumber\\
  &&X_2=R\frac{\sin \tau}{\cos \sigma}=-R\frac{t}{z}\nonumber\\
&&X_{3}=R\tan \sigma =-\frac{z}{2}(1-\frac{R^2+t^2}{z^2}).\label{lower}
\end{eqnarray}
 We can immediately verify that
 \begin{eqnarray}
-X_{1}^2-X_2^2+X_3^2=-R^2.\label{ads2L}
 \end{eqnarray}
 Thus, anti-de Sitter spacetime ${\rm AdS}^2$ is the one-sheeted hyperboloid embedded in $\mathbb{R}^{2,1}$ with negative constant curvature.
 
 It is easily seen that the group $SO(2,1)=SU(1,1)/\mathbb{Z}_2$ is the group of isometries (translations, dilatations and special conformal transformations) of ${\rm AdS}^2$ in the same way that the group $SO(3)=SU(2)/\mathbb{Z}_2$ is the group of isometries (rotations) of $\mathbb{S}^2$.

 For ${\rm AdS}^{2}$ the ranges of the cylindrical coordinates are $\sigma\in ]-\pi/2,\pi/2[$ and  $\tau\in [0,2\pi]$ (which becomes $[-\infty,+\infty]$ as we unwrap to the universal cover by lifting the restriction on the range of the time coordinate in order to avoid closed timelike curves). Thus, ${\rm AdS}^{2}$ is a rectangle with bases at $\tau=\pm \infty$, a center at $\sigma=0$ and two spatial infinities (boundaries) at $\sigma\longrightarrow \pm \pi/2$.

The higher-dimensional generalization of equation (\ref{lower}) gives the global system of coordinates  $(\tau, \sigma,\hat{x}_i)$ on ${\rm AdS}^{d+1}$. Explicitly, we have
\begin{eqnarray}
&&X_{1}=R\frac{\cos \tau}{\cos \sigma}=-\frac{z}{2}(1+\frac{R^2+\vec{x}^2-t^2}{z^2})\nonumber\\
  &&X_2=R\frac{\sin \tau}{\cos \sigma}=-R\frac{t}{z}\nonumber\\
&&X_i=R\tan \sigma \hat{x}_i=-R\frac{x_i}{z}~,~2<i<d+2\nonumber\\
&&X_{d+2}=R\tan \sigma \hat{x}_{d+2}=-\frac{z}{2}(1-\frac{R^2-\vec{x}^2+t^2}{z^2}).\label{p}
\end{eqnarray}
 Thus, for ${\rm AdS}^{d+1}$ with dimension $d+1>2$ the range of the cylindrical coordinate $\sigma$  is instead given by $\sigma\in [0,\pi/2[$ (due to the extra angular variable $\hat{x}_i$ which describes a sphere $\mathbb{S}^{d-1}$, i.e.  $\sum_i\hat{x}_i^2=1$). Thus, ${\rm AdS}^{d+1}$  is a cylinder with bases at $\tau=\pm \infty$, a center at $\sigma=0$, a single spatial infinity at $\sigma\longrightarrow \pi/2$, while going around the cylinder corresponds to the angular variable $\hat{x}_i$.

    The fact that we have for ${\rm AdS}^2$ two disconnected one-dimensional boundaries at $\sigma=\pm \pi/2$  makes this case very different from higher dimensional anti-de Sitter spacetimes and is probably what makes the ${\rm AdS}^2$/${\rm CFT}_1$ correspondence the most mysterious case among all examples of the AdS/CFT correspondence. For example, see \cite{Strominger:1998yg,Spradlin:1999bn} and \cite{Cadoni:1998sg,Cadoni:1999ja}.
     
    The (Euclidean) ${\rm AdS}_2$  is the hyperbolic space $\mathbb{H}^2$  (a pseudosphere) embedded in $\mathbb{R}^{1,2}$ with metric $\eta=(-1,+1,+1)$ which is obtained by the Wick rotation $X_2\longrightarrow -iX_2$ and $\tau\longrightarrow -i\tau$. The isometry group is $SO(1,2)$ and therefore all group theory considerations remain virtually unchanged.

    The metric of Euclidean ${\rm AdS}_2$ in the so-called Poincar\'e coordinates $(t,z)$, defined also by equation (\ref{lower}),  is given by 

 \begin{eqnarray}
   ds^2&=&\frac{R^2}{z^2}(dz^2+dt^2)\nonumber\\
   &=& R^2\big(\frac{du^2}{u^2}+u^2dt^2)~,~u=\frac{1}{z}.\label{met2}
 \end{eqnarray}
 The Poincar\'e patch is perhaps the most important system of coordinates on ${\rm AdS}$ spacetime. The point $u=\infty$ or equivalently the point $z=0$ is the conformal boundary (the spatial infinities $\sigma\longrightarrow \pm\pi/2$) whereas the point $u=0$ or equivalently $z=\infty$ is the horizon which is a single point in the Euclidean case. In the Minkowski case the Poincar\'e coordinates are called a patch because they cover only half of the spacetime which obviously can be extended beyond the horizon.

 In the case of a higher dimensional anti-de Sitter spacetime ${\rm AdS}^{d+1}$ the conformal boundary $z=0$ is  a Minkowski spacetime $\mathbb{M}^{d}$ with metric $\eta=(-1,+1,...,+1)$.  Indeed, we can think of $u$ as the energy scale of the conformal field theory living on the Minkowski spacetime $\mathbb{M}^{d}$ which is foliated at $z=1/u$. Alternatively, $z$ and $u$ can be thought of as a lattice spacing and an ultraviolet cutoff respectively after Euclidean continuation. 

In the case of  Euclidean  ${\rm AdS}_2$ the global coordinates are given explicitly by the following relations

    \begin{eqnarray}
&&X_{1}=R\frac{\cosh \tau}{\cos \sigma}=-\frac{z}{2}(1+\frac{R^2+t^2}{z^2})\nonumber\\
  &&X_2=R\frac{\sinh \tau}{\cos \sigma}=-R\frac{t}{z}\nonumber\\
&&X_{3}=R\tan \sigma =-\frac{z}{2}(1-\frac{R^2-t^2}{z^2}).\label{global}
\end{eqnarray}
The Euclidean  ${\rm AdS}_2$ is then given by the pseudosphere
 
\begin{eqnarray}
-X_1^2+X_2^2+X_{3}^2=-R^2.\label{ads2E}
\end{eqnarray}
The isometry group is still given by $SO(1,2)=SU(1,1)/\mathbb{Z}_2$.

\section{The noncommutative ${\rm AdS}^2_{\theta}$}

All Poisson manifolds are characterized by a symplectic two-form which can be quantized in the usual way. A celebrated example is the fuzzy sphere \cite{Hoppe,Madore:1991bw}. The noncommutative  ${\rm AdS}^2_{\theta}$ is in fact a very close relative of the fuzzy sphere  \cite{Ho:2000fy,Ho:2000br,Jurman:2013ota,Pinzul:2017wch}. Indeed, in Euclidean signature the noncommutative ${\rm AdS}^2_{\theta}$ is a pseudo-sphere embedded in Minkowski spacetime $\mathbb{M}^{1,2}$. As a consequence, the symmetry group underling the fuzzy sphere is the rotation group $SO(3)=SU(2)/\mathbb{Z}_2$ whereas the symmetry group underlying the noncommutative  ${\rm AdS}^2_{\theta}$  is the Lorentz group $SO(1,2)=SU(1,1)/\mathbb{Z}_2$. The Lorentz group $SO(1,2)$ is a subgroup of the $d=1$ conformal group $SL(2,R)$.  Thus, familiarity with the representation theory of the Lorentz group $SO(1,2)$ is important. See for example \cite{barg,bns} and \cite{Mukunda:1974gb,Girelli:2015ija,Basu:1981ju}.

The noncommutative geometry of ${\rm AdS}^2_{\theta}$ is defined in terms of a spectral triple $({\cal A}, {\cal H},\Delta)$ consisting of an algebra ${\cal A}$, a Hilbert space ${\cal H}$ and a Laplacian $\Delta$  \cite{Connes:1996gi}. In here it will be sufficient to specify the geometry in terms of an appropriate set of coordinate coperators with a suitable commutative or semi-classical limit.

Euclidean anti-de Sitter space ${\rm AdS}^2$  is the co-adjoint orbit $SO(1,2)/SO(2)$ which admits a  non-degenerate symplectic two-form $\omega$ given by
\begin{eqnarray}
  \omega=\frac{1}{\kappa}\frac{R}{\cos^2\sigma} d\tau\wedge d\sigma.
\end{eqnarray}
This two-form is closed, i.e. $d\omega=0$ and its quantization (which leads to the quantization of the deformation parameter $\kappa$) yields the noncommutative anti-de Sitter space ${\rm AdS}^2_{\theta}$. Indeed, this two-form can be inverted (since it is non-degenerate) allowing us to define a Poisson structure which can  be quantized in the usual way by replacing the Poisson brackets with commutators. The Poisson structure on anti-de Sitter space is given explicitly (in the coordiantes $\sigma$ and $\tau$) by

\begin{eqnarray}
\{f,g\}=\kappa\frac{\cos^2\sigma}{R}(\partial_{\tau}f\partial_{\sigma}g-\partial_{\sigma}f\partial_{\tau}g).
\end{eqnarray}
We compute immediately the fundamental Poisson brackets

\begin{eqnarray}
  \{X^a,X^b\}=\kappa f^{ab}~_cX^c. 
\end{eqnarray}
The structure constants are given by $f^{ab}~c=\epsilon^{ab}~c$ for Euclidean ${\rm AdS}^2$. Canonical quantization proceeds now in the usual way by introducing a Hilbert space ${\cal H}$, replacing the coordinate functions $X^a$ with coordinate operators $\hat{X}^a$ and replacing the fundamental Poisson brackets $\{.,.\}$ with the fundamental commutators $[.,.]/i$. We obtain then the commutation relations

\begin{eqnarray}
  [\hat{X}^a,\hat{X}^b]=i\kappa f^{ab}~_c\hat{X}^c. \label{45}
\end{eqnarray}
The noncommutative ${\rm AdS}^2_{\theta}$ is also given by the embedding relation (which generalizes (\ref{ads2E}))

\begin{eqnarray}
  -\hat{X}_1^2+\hat{X}_2^2+\hat{X}_{3}^2=-R^2.\label{46}
\end{eqnarray}
More explicitly, the coordinate operators $\hat{X}^a$ are given by

\begin{eqnarray}
    \hat{X}^a=\kappa K^a.\label{47}
  \end{eqnarray}
The $K^a$ are the generators of the Lie group $SO(1,2)$ in the irreducible representations given by the discrete series $D_k^{\pm}$ with $k=\{1/2,1,2/3,2,3/2,...\}$. These  are infinite dimensional unitary irreducible representations corresponding to the lowest  and highest weight states given respectively by the Hilbert spaces

  \begin{eqnarray}
      {\cal H}_k^+=\{|km\rangle; m=k,k+1,k+2,...\}.
    \end{eqnarray}

    \begin{eqnarray}
      {\cal H}_k^-=\{|km\rangle; m=-k,-(k+1),-(k+2),...\}.
    \end{eqnarray}
 The Casimir is given by $C=-K_1^2+K_2^2+K_3^2=-k(k-1).{\bf 1}$ whereas the action of the generators $K^1$ and  $K^{\pm}=-K^3\pm i K^2$ is given by

\begin{eqnarray}
  &&K^1|km\rangle =m|km\rangle\nonumber\\
  &&K^+|km\rangle=\sqrt{m(m+1)-k(k-1)}|km+1\rangle\nonumber\\
  &&K^-|km\rangle=\sqrt{m(m-1)-k(k-1)}|km-1\rangle.
\end{eqnarray}
The   integer $k$ must be quantized in such a way that we have the relation \cite{Pinzul:2017wch}

  \begin{eqnarray}
    \frac{R^2}{\kappa^2}=k(k-1).
  \end{eqnarray}
 The commutative limit $\kappa\longrightarrow 0$ corresponds therefore to $k\longrightarrow\infty$.

The radius operator is obviously defined by 

\begin{eqnarray}
      &&\hat{u}=\frac{1}{\hat{z}}=\frac{\hat{X}_1-\hat{X}_3}{R^2}.
    \end{eqnarray}
We compute immediately the expectation value $\langle km|\hat{u}|km\rangle=\kappa m/R^2$ which approaches $\pm \infty$ as $m\longrightarrow \pm\infty$ corresponding to the two representations $D_k^{\pm}$ \cite{Pinzul:2017wch}. These two limits $m\longrightarrow \pm\infty$ correspond to the two boundaries of noncommutative ${\rm AdS}^2_{\theta}$.

 \section{The $SO(1,2)-$coherent states}
 In this section (which can be skipped on a first read) we will present the $SO(1,2)-$coherent state which will allow us to intuitively visualize noncommutative ${\rm AdS}^2_{\theta}$. The over-complete basis of coherent states provides also a natural starting point for the construction of the Weyl map and the corresponding Moyal-Weyl star product (using also Darboux theorem) and hence allows us a concrete handle on the commutative limit.

    Another very useful system of coordinates on  Euclidean ${\rm AdS}^{2}$ or $\mathbb{H}^2$ is the pseudosphere coordinates and their Poincare disk (complex) coordinates which should be thought of as the corresponding stereographic  projection.

Recall the spherical coordinates $\theta$ and $\phi$ on the sphere $\mathbb{S}^2$ and their stereographic projection (complex) coordinate $\xi$ given respectively by (with $n_i=X_i/R$)
    \begin{eqnarray}
      n_1=\sin\theta\cos\phi~,~n_2=\sin\theta\sin\phi~,~n_3=\cos\theta.
    \end{eqnarray}
    \begin{eqnarray}
      \xi=-\frac{n_1-in_2}{1+n_3}=-\tan\frac{\theta}{2}\exp(-i\phi).
    \end{eqnarray}
    Similarly, the pesudosphere coordinates $\tau$ and $\phi$ and the Poincare disk (complex) coordinates $\xi$ on   ${\rm AdS}^{2}$ are given by (with $n_i=X_{i+1}/R$ and $n_0>0$)

 \begin{eqnarray}
      &&n_0=\cosh\tau~,~n_1=\sinh\tau\cos\phi~,~n_2=\sinh\tau\sin\phi.\nonumber\\
    \end{eqnarray}
    \begin{eqnarray}
      \xi=\frac{n_1-in_2}{1+n_0}=\tanh\frac{\tau}{2}\exp(-i\phi).
    \end{eqnarray}
    We compute the metric
    \begin{eqnarray}
      ds^2=d\tau^2+\sinh^2\tau d\phi^2=\frac{4}{(1-|\xi|^2)^2}(d\xi_1^2+d\xi_2^2).
    \end{eqnarray}
    From this we obtain the measure (with $\xi=\rho\exp(i\kappa)$)
    \begin{eqnarray}
      \sinh\tau d\tau d\phi=\frac{4}{(1-|\xi|^2)^2}d\xi_1d\xi_2=\frac{4\rho}{(1-\rho^2)^2}d\rho d\kappa.
    \end{eqnarray}
The upper sheet of the hyperboloid $\mathbb{H}^2$ ($n_0\geq 0$) which is precisely Euclidean ${\rm AdS}^2$ is then isomorphic to the disk $|\xi|\leq 1$.
    
    Coherent states are quantum states in the Hilbert space with the most classical properties. They can be built in a straightforward way for coadjoint orbits such as $\mathbb{S}^2=SU(2)/U(1)$ and ${\rm AdS}^2=SU(1,1)/U(1)$ in which there is an intimate relation between the geometry of the underlying manifold and the geometry of the isometry group acting on it \cite{perel}.  In this case the coherent states represent in a precise sense the points of the quantized geometry.

    Thus, starting from any point $\vec{n}_0$ in the manifold we can reach any other point $\vec{n}$ (since the action of the group is transitive) through the application of an element $g$ of the group $G=SU(2)$ (for the sphere) and $G=SU(1,1)$ (for the pseudosphere), i.e. $\vec{n}=g\vec{n}_0$. The point $\vec{n}_0$ is associated with a state $|\psi_0\rangle$ of minimum uncertainty in the appropriate irreducible representation of the group. For example, $|\psi_0\rangle=|j,-j\rangle$ for the sphere and $|\psi_0\rangle=|k,k\rangle$ for the pseudosphere, i.e. $\vec{n}_0$ corresponds to the north pole $(0,0,1)$ on the sphere and to the "north pole" $(1,0,0)$ on the pseudosphere. The state $|\psi\rangle$ associated with the rotated point $\vec{n}$ is then given by $U(g)|\psi_0\rangle$ where $U(g)$ is the representation of the group element $g$ in the considered representation. We write
    \begin{eqnarray}
      &&|\psi_0\rangle=|\vec{n}_0\rangle\nonumber\\
&&|\psi\rangle=U(g)|\psi_0\rangle\equiv |\vec{n}\rangle.
    \end{eqnarray}
    The most important case for us here is the case of the non-compact group $SU(1,1)$ in the infinite dimensional irreducible representations given by the positive discrete series $D^+_k$. These  are lowest state representations characterized by an integer Bargmann index $k$. The Lie algebra is given explicitly (with $K^1=L_0$, $K^2=L_1$, $K^3=L_2$)
\begin{eqnarray}
[L_0,L_1]=iL_2~,~[L_0,L_2]=-iL_1~,~[L_1,L_2]=-iL_0.
    \end{eqnarray}
Equivalently (with $L_{\pm}=\pm iK^{\pm}$)
\begin{eqnarray}
[L_0,L_{\pm}]=\pm L_{\pm}~,~[L_+,L_-]=-2L_0.
\end{eqnarray}
The Casimir is given by
\begin{eqnarray}
C=-L_0^2+L_1^2+L_2^2\equiv -\hat{C}.
\end{eqnarray}
The states of the irreducible representation $D^+_k$ are given by (with $~k=1,\frac{3}{2},2,\frac{5}{2},...$) 
\begin{eqnarray}
  &&\hat{C}|k,m\rangle=k(k-1)|k,m\rangle\nonumber\\
&&L_0|k,m\rangle=(k+m)|k,m\rangle~,~m\geq 0.
\end{eqnarray}
The group $SU(1,1)$ is the set of all $2\times 2$ matrices $g$ with unit determinant which satisfy $gJg^{\dagger}=J$ where $J={\rm diag}(1,-1)$. For $SU(2)$, $J={\rm diag}(1,1)$. Thus, $SU(2)$ is the set of all   $2\times 2$ matrices $g$ with unit determinant which leave the form $|z_1|^2+|z_2|^2$ invariant whereas $SU(1,1)$  is the set of all $2\times 2$ matrices $g$ with unit determinant which leave the form $|z_1|^2-|z_2|^2$ invariant. We can bring $g\in SU(1,1)$ into the form
\begin{eqnarray}
 g= \left( \begin{array}{cc}
\alpha & \beta \\
\bar{\beta} & \bar{\alpha} \end{array} \right)~,~{\rm det} g=|\alpha|^2-|\beta|^2=1.
  \end{eqnarray}
The anti-de Sitter spacetime   ${\rm AdS}^2$ is the coadjoint orbit $SU(1,1)/U(1)$ which means that points $\vec{n}$ of the AdS spacetime are actually equivalence classes $[gh]$ where $g\in SU(1,1)$ and $h$ is an element of the stability group $U(1)$ given explicitly by
\begin{eqnarray}
 h= \left( \begin{array}{cc}
\alpha & 0 \\
0 & \bar{\alpha} \end{array} \right).
\end{eqnarray}
This can be seen explicitly from the relation
\begin{eqnarray}
  gL_0g^{-1}=-L_an^a~,~g^{-1}=Jg^{\dagger}J.
\end{eqnarray}
The $L_0$ is the generator of the stability group $U(1)$ and it is seen that its orbit given by the coadjoint action $gL_0g^{-1}$ for some $g\in SU(1,1)$ and any $h\in U(1)$ is labeled by the point $\vec{n}$ of   ${\rm AdS}^2$. Indeed, this orbit is spanned under the right action $g\longrightarrow gh$ and since $hL_0h^{-1}=L_0$ the point $\vec {n}$ of ${\rm AdS}^2$ can be associated uniquely to the class $[gh]$ of $SU(1,1)/U(1)$.

In a general irreducible representation of the group $SU(1,1)$ corresponding to a discrete series with Bargmann index $k$ the above equation reads
\begin{eqnarray}
  U(g)L_0U(g^{-1})=-L_an^a.
\end{eqnarray}
Thus, if the fiducial coherent state $|\psi_0\rangle=|k,k\rangle$ is an eigenstate of $L_0$ with eigenvalue $k$ then the rotated coherent state $|\psi\rangle=U(g)|\psi_0\rangle$ is an eigenstate of  $-L_an^a$ with the same eigenvalue $k$, i.e. $-L_an^a$ is a rotated $L_0$. We have then
\begin{eqnarray}
L_0|\psi_0\rangle=k|\psi_0\rangle\Rightarrow  -L_an^a|\psi\rangle=k|\psi\rangle.
\end{eqnarray}
Similarly, since the fiducial vector satisfies $L_-|\psi_0 \rangle=0$ (because it is a lowest state vector) the rotated coherent state must satisfy $ U(g)L_-U(g^{-1})|\psi\rangle=0$.

Similar considerations hold on the sphere $\mathbb{S}^2$ which is the coadjoint orbit $SU(2)/U(1)$. Thus, points $ \vec{n}$ of the sphere  $\mathbb{S}^2$ are associated uniquely to the equivalence classes $[gh]$ of  $SU(2)/U(1)$ via the relation $g\sigma_3g^{-1}=\sigma_an_a$.

In the fundamental  representation of $SU(1,1)$ (which is non-unitary) we have $L_0=\sigma_3/2$, $L_{1}=i\sigma_1/2$ and $L_2=i\sigma_2/2$ and hence the above relation takes the form (noting also that $J\equiv \sigma_3$)
\begin{eqnarray}
gg^{\dagger}\sigma_3=\sigma_3n_0-i\sigma_1n_1-i\sigma_2.
\end{eqnarray}
By choosing $h$ appropriately we can choose the representative element $g_{\vec{n}}$ of the equivalence class $[gh]$ to be of the form (with $\alpha>0$ corresponding to the upper sheet of the hyperboloid)
\begin{eqnarray}
 g_{\vec{n}}= \left( \begin{array}{cc}
\alpha & \beta \\
\bar{\beta} & {\alpha} \end{array} \right).
  \end{eqnarray}
By substituting in the previous equation we obtain immediately
\begin{eqnarray}
n_0=\alpha^2+|\beta|^2~,~n_1=2\alpha \beta_2~,~n_2=2\alpha\beta_1.
\end{eqnarray}
After some algebra we find the solution
\begin{eqnarray}
\alpha=\cosh\frac{\tau}{2}~,~\beta=i\sinh\frac{\tau}{2}e^{-i\phi}.
\end{eqnarray}
Hence the representative element $g_{\vec{n}}$ of the coset $[gh]$ is parametrized by the local coordinates $\tau$ and $\phi$ on the AdS spacetime as follows  (with $m_1=\sin\phi$, $m_2=-\cos\phi$ and $\zeta=-i\tau(m_1-im_2)/2$)
\begin{eqnarray}
 g_{\vec{n}}&=& \left( \begin{array}{cc}
\cosh\frac{\tau}{2} & i\sinh\frac{\tau}{2}e^{-i\phi} \\
-i\sinh\frac{\tau}{2}e^{i\phi} & \cosh\frac{\tau}{2} \end{array} \right)\nonumber\\
&=&\exp(\tau(m_1L_1+m_2L_2))\nonumber\\
 &=&\exp(\zeta L_+-\bar{\zeta}L_-).
\end{eqnarray}
Similar considerations on $\mathbb{S}^2=SU(2)/U(1)$ will lead to the group element (with $m_1=\sin\phi$, $m_2=-\cos\phi$ and $\zeta=i\theta(m_1-im_2)/2$)
\begin{eqnarray}
 g_{\vec{n}}&=& \left( \begin{array}{cc}
\cos\frac{\theta}{2} & -\sin\frac{\theta}{2}e^{-i\phi} \\
\sin\frac{\theta}{2}e^{i\phi} & \cos\frac{\theta}{2} \end{array} \right)\nonumber\\
&=&\exp(i \theta (m_1L_1+m_2L_2))\nonumber\\
  &=&\exp(\zeta L_+-\bar{\zeta}L_-).
\end{eqnarray}
The group $SU(1,1)$ is a real form of the complex group $SL(2,\mathbb{C})$. Thus, any element $g_{\vec{n}}$ of  $SU(1,1)$ understood as an element of the complex group $SL(2,\mathbb{C})$ admits the so-called Gaussian decomposition, viz
\begin{eqnarray}
 g_{\vec{n}}= \left( \begin{array}{cc}
\alpha & \beta \\
\gamma & \delta \end{array} \right)=z_+hz_-.
\end{eqnarray}
\begin{eqnarray}
 z_+= \left( \begin{array}{cc}
1 & {\beta}/{\delta} \\
0 & 1 \end{array} \right)~,~h= \left( \begin{array}{cc}
{1}/{\delta} & 0 \\
0 & \delta \end{array} \right)~,~z_-= \left( \begin{array}{cc}
1 & 0 \\
{\gamma}/{\delta} & 1 \end{array} \right).\nonumber\\
\end{eqnarray}
For our case, we compute immediately
\begin{eqnarray}
z_+&=& \left( \begin{array}{cc}
1 & i\tanh\frac{\tau}{2}e^{-i\phi} \\
0 & 1 \end{array} \right)=1+\tanh\frac{\tau}{2}e^{-i\phi}L_+\nonumber\\
&=&\exp(\xi L_+)\nonumber\\
h&=& \left( \begin{array}{cc}
{1}/\cosh\frac{\tau}{2} & 0 \\
0 & \cosh\frac{\tau}{2} \end{array} \right)\nonumber\\
&=&\exp(-\eta K_0)\nonumber\\
  z_-&=& \left( \begin{array}{cc}
1 & 0 \\
-i\tanh\frac{\tau}{2}e^{i\phi} & 1 \end{array} \right)=1+\tanh\frac{\tau}{2}e^{i\phi}L_-\nonumber\\
&=&\exp(\bar{\xi} L_-).
\end{eqnarray}
In the second equation above $\eta$ is given by
\begin{eqnarray}
\eta=2\ln\cosh \frac{\tau}{2}=-\ln(1-|\xi|^2).
\end{eqnarray}
The representative element $g_{\vec{n}}$ of the coset $[gh]$ associated with the point $\vec{n}$ of ${\rm AdS}^2$  is given then explicitly by
\begin{eqnarray}
  g_{\vec{n}}=\exp(\xi L_+)\exp(-\eta L_0)\exp(\bar{\xi} L_-).
\end{eqnarray}
Although this result is obtained in the fundamental representation of the group $SU(1,1)$ (since the generators $L$'s were assumed to be  proportional to the Pauli matrices) it is in fact valid (since the parameters are independent of the representation) in any irreducible representation of the group corresponding to a discrete series with Bargmann index $k$. We have then the so-called displacement operator $U^{(k)}(g_{\vec{n}})$ associated with the element $g_{\vec{n}}$ of the coset $[gh]$ or equivalently with the point $\vec{n}$ of ${\rm AdS}^2$ given explicitly by
\begin{eqnarray}
  U^{(k)}(g_{\vec{n}})&=&\exp(\zeta L_+-\bar{\zeta}L_-)\nonumber\\
  &=&\exp(\xi L_+)\exp(-\eta L_0)\exp(\bar{\xi} L_-).
\end{eqnarray}
We can immediately compute the coherent state (using also $L_-|k,k\rangle=0$ and $L_0|k,k\rangle=k|k,k\rangle$)
\begin{eqnarray}
  |\vec{n}\rangle&=&U^{(k)}(g_{\vec{n}})|\vec{n}_0\rangle\nonumber\\
  &=&\exp(\xi L_+)\exp(-\eta L_0)\exp(\bar{\xi} L_-)|k,k\rangle\nonumber\\
  &=&(1-|\xi|^2)^k\sum_{n=0}^{\infty}\sqrt{\frac{\Gamma(n+2k)}{n!\Gamma(2k)}}\xi^n|k,k+n\rangle\nonumber\\
  &\equiv &|\xi\rangle.
\end{eqnarray}
In the above equation we have also used the relation
\begin{eqnarray}
  |k,k+n\rangle
    &=&\sum_{n=0}^{\infty}\sqrt{\frac{\Gamma(2k)}{n!\Gamma(n+2k)}}(L_+)^n|k,k\rangle.
\end{eqnarray}
We compute the resolution of unity
\begin{eqnarray}
\int d\mu_k(\xi)|\xi\rangle\langle\xi|=\sum_{n=0}^{\infty}|k,k+n\rangle\langle k,k+n|=1.
\end{eqnarray}
In deriving this equation we use also the following result
\begin{eqnarray}
\int d\mu_k(\xi)\xi^n\bar{\xi}^m(1-|\xi|^2)^{2k}=\frac{n!\Gamma(2k)}{\Gamma(n+2k)}\delta_{n,m}.
\end{eqnarray}
The measure is then given explicitly by
\begin{eqnarray}
  d\mu_k(\xi)=\frac{2k-1}{\pi}\frac{d\xi_1d\xi_2}{(1-|\xi|^2)^2}.
\end{eqnarray}
Similarly, we compute the inner product between any two coherent states to be given by 
\begin{eqnarray}
\langle \xi_1|\xi_2\rangle=\frac{(1-|\xi_1|^2)^k(1-|\xi_2|^2)^k}{(1-\bar{\xi}_1\xi_2)^{2k}}.
\end{eqnarray}
We conclude with the very important result 
\begin{eqnarray}
  \langle \xi|L_a|\xi\rangle=k n_a.
\end{eqnarray}
This shows explicitly that the coordinate operators on the noncommutative  ${\rm AdS}^2_{\theta}$ are given precisely by the operators $L_a/k$ which were obtained previously by a direct quantization of the symplectic structure on  ${\rm AdS}^2$.

\section{Yang-Mills matrix models, UV-IR mixing and phase structure}
We can think of "noncommutative geometry" as the correct theory of "classical gravity" (first quantization) while the corresponding "matrix models" we can think of them as the corresponding theory of "quantum gravity" (second quantization).

For example, it can be shown that Yang-Mills matrix models captures the noncommutative geometry of the fuzzy sphere $\mathbb{S}^2_N$ as well as the corresponding geometrical/gravitational quantum fluctuations about this background \cite{Alekseev:2000fd,Iso:2001mg,Steinacker:2003sd}. These Yang-Mills matrix models are essentially truncation of the IKKT matrix model \cite{Ishibashi:1996xs} to lower dimensions with and without supersymmetric mass deformations. This is a typical example of emergent geometry and emergent gravity where the geometrical/gravitational degrees of freedom are mostly encoded in a gauge theory.

Another example is the noncommutative ${\rm AdS}^2_{\theta}$ in Euclidean signature which is given by equations (\ref{45}) and (\ref{46}). This space is really a quantized pseudo-sphere, i.e. a quantized Hyperboloic space $\mathbb{H}^2$. It is not difficult to convince ourselves that one must start in this case from an IKKT-type model given by the Yang-Mills matrix model
\begin{eqnarray}
S[D]&=&N {\rm Tr}(-\frac{1}{4}[D_a,D_b][D^a,D^b]+\frac{2i}{3}\kappa f_{abc}D^aD^bD^c\nonumber\\
&+&\beta D_aD^a).\label{YM}
\end{eqnarray}
The quadratic term is added for completeness. The cubic term is effectively Myers effect \cite{Myers:1999ps} which is responsible here for the condensation of matrix/noncommutative geometry into classical/commutative geometry. The ambient metric is $\eta=(-1,+1,+1)$, $D_a=(D_a)^{\dagger}$ are three matrices in ${\rm Mat}(\infty,\mathbb{C})$ and $f_{abc}$ are the structure constants of $SO(1,2)$. Hence this model is invariant under $SO(1,2)$ rotations as well as under gauge transformations $D_a\longrightarrow UD_aU^{\dagger}$ and under  translations $D_a\longrightarrow D_a+c$. 

At the level of the matrix model the two main differences between the noncommutative ${\rm AdS}^2_{\theta}$ case and the fuzzy sphere case are:
\begin{enumerate}
\item The trace ${\rm Tr}$ is infinite dimensional.
\item The action lives in an ambient space with Lorentzian signature.
\end{enumerate}
Regarding the second difference we can first note that in the geometric ${\rm AdS}^2$ phase the Yang-Mills term of the matrix model is positive definite despite the ambient Lorentzian signature. This will be discussed further below.

Furthermore, since it is really hard to see how does the matrix model (\ref{YM}) know about the ambient Lorentzian signature, it seems that this matrix model actually reproduces the two spaces: 
\begin{eqnarray}
 -\hat{X}_1^2+\hat{X}_2^2+\hat{X}_{3}^2=-R^2~,~\mathbb{R}^3.\label{mid1}
\end{eqnarray}
\begin{eqnarray}
 -\hat{X}_1^2+\hat{X}_2^2+\hat{X}_{3}^2=-R^2~,~\mathbb{M}^{1,2}.\label{mid2}
\end{eqnarray}
The first space is a two-sheeted hyperboloid with a non-constant curvature, some of its Killing vectors fail to leave the ambient metric invariant \cite{Bengtsson}, whereas the second space is the  pseudo-sphere or the hyperboloic space $\mathbb{H}^2$, i.e.   ${\rm AdS}^2$ in Euclidean signature which is a space of constant curvature. The first space is seen to lie midway between the sphere and the pseudo-sphere and it is quantized by means of the non-unitary finite irreducible representations $F_k$ of $SO(1,2)$ as opposed to the discrete series $D_k^{\pm}$ used to quantize the pseudo-sphere  ${\rm AdS}^2$.

The first difference between ${\rm AdS}^2_{\theta}$ and $\mathbb{S}^2_N$ is equally important and equally serious. Indeed, the infinite dimensionality of the matrix algebra (and the fact that the operator Laplacian has the same spectrum as the commutative one \cite{ydri}) makes the noncommutative ${\rm AdS}^2$ quite similar to its commutative counterpart in stark contrast with the case of the fuzzy sphere $\mathbb{S}^2_N$.

In the following we need to regularize the matrix model (\ref{YM}) by cutting off the trace ${\rm Tr}$ in such a way that only $N_k=2k-1$ states in the Hilbert space ${\cal H}_k^{\pm}$ are included where $k$ is the pseudo-spin quantum number associated with the irreducible representation $D_k^{\pm}$ of $SO(1,2)$. Thus, ${\rm Tr}={\rm Tr}_k$ where ${\rm Tr}_k{\bf 1}=N_k$. In analogy with the fuzzy sphere case the normalization ${N}$ in front of the action (\ref{YM}) will be chosen to be precisely equal to $N_k$. We employ then in (\ref{YM}) the regularization
\begin{eqnarray}
&&{\rm Tr}\longrightarrow {\rm Tr}_k\nonumber\\
&&N\longrightarrow N_k=2k-1.
\end{eqnarray}
The variation of the action and the equation of motion are given by

\begin{eqnarray}
&&\delta S=-N{\rm Tr}\delta D^a\bigg([D^b,F_{ab}]-2\beta D_a\bigg)\equiv 0\nonumber\\
&&\Rightarrow [D^b,F_{ab}]-2\beta D_a=0.\label{eom}
\end{eqnarray}
The field strength $F_{ab}$ of the gauge field $D_a$ is given by
\begin{eqnarray}
F_{ab}=[D_a,D_b]-i\kappa f_{abc}D^c.\label{eom}
\end{eqnarray}
A solution of these equations of motion is given by

\begin{eqnarray}
D^a=\phi K^a~,~\phi=\kappa\varphi.\label{sp0}
\end{eqnarray}
The order parameter $\varphi$ is given by (using among other things $\epsilon^{abc}\epsilon_{abd}=-2\delta^c_d$ and $\tau= -\beta/\kappa^2$)
\begin{eqnarray}
\varphi^3-\varphi^2+\tau\varphi=0\Rightarrow \varphi_0=0~,~\varphi_{\pm}=\frac{1}{2}\big(1\pm\sqrt{1-4\tau}\big).\label{sp1}\nonumber\\
\end{eqnarray}
The classical potential, i.e. the value of the action in this configuration $\phi$ is given by (with $C=K_aK^a=-k(k-1).{\bf 1}$ and $k(k-1)=R^2/\kappa^2$)
\begin{eqnarray}
V&=&N_k{\rm Tr} K_aK^a\bigg[-\frac{1}{2}\phi^4+\frac{2}{3}\kappa\phi^3+\beta\phi^2\bigg]\nonumber\\
&=&2N_k^2\kappa^2 R^2\bigg[\frac{1}{4}\varphi^4-\frac{1}{3}\varphi^3+\frac{1}{2}\tau\varphi^2\bigg]\nonumber\\
&=&2N_k^2\kappa^4 k(k-1)\bigg[\frac{1}{4}\varphi^4-\frac{1}{3}\varphi^3+\frac{1}{2}\tau\varphi^2\bigg].\label{sp2}
\end{eqnarray}
For stability (the sign of the quartic term needs to be positive) the Casimir operator $C=K_aK^a$ is required to be negative which selects the discrete representations $D_k^{\pm}$ of $SO(1,2)$ (used for ${\rm AdS}^2$) as well as the finite representations $F_k$ (which are non-unitary and can be used to define a matrix approximation of (\ref{mid1})). Thus, in the second line in the above equation we have assumed that $K^a$ are the generators of $SO(1,2)$ in the irreducible representations given by the discrete series $D_k^{\pm}$. We can also consider the case of reducible representations of $SO(1,2)$ consisting of direct sum of different discrete series.

Before we proceed to quantum corrections two quick but important remarks are in order. First, the two results (\ref{sp1}) and (\ref{sp2}) are identical to those obtained in the fuzzy sphere case  \cite{Delgadillo-Blando:2012skh} and as a consequence similarity between the phase diagrams ensues. Second, the conditions (\ref{45}) and (\ref{46}) are  solved by (\ref{sp0})  generating effectively both spaces (\ref{mid1}) and (\ref{mid2}). We can think of (\ref{mid1}) as a non-unitary realization of noncommutative ${\rm AdS}^2$.

The calculation of the one-loop corrections in the case of the noncommutative anti-de Sitter space ${\rm AdS}^2_{\theta}$ proceeds in  exactly the same way as the analogous calculation in the case of the fuzzy sphere $\mathbb{S}^2_N$ \cite{CastroVillarreal:2004vh}. The only difference here lies in the signature of the metric. Furthermore, these IKKT-type Yang-Mills models, such as (\ref{YM}), are generally characterized by one-loop dominance, i.e. quantum corrections are dominated by the one-loop result which makes this calculation all the more fundamental. For example, this one-loop dominance is studied in the case of the fuzzy sphere matrix model in \cite{Azuma:2004zq}.

Quantization around the ${\rm AdS}^2_{\theta}$ background (\ref{sp0}), using the background field method gives, after gauge fixing in the Lorentz gauge, the effective action
 \begin{eqnarray}
{\Gamma}&=&S+\frac{1}{2}{\rm tr}\log{\Omega}-{\rm tr}\log{\cal D}^2.
\end{eqnarray}
The Laplacian operator ${\Omega}$ on the space of field configurations is given explicitly by the formula 
\begin{eqnarray}
{\Omega}_{ab}&=& {\cal D
}_c^2{\delta}_{ab}-2{\cal
F}_{ab}+2\beta{\delta}_{ab}.
\end{eqnarray}
The first term is due to the gauge field while the second term is due to the ghost field. The notation ${\cal D}_a$ and ${\cal F}_{ab}$ means that the covariant derivative $D_a$ and the curvature $F_{ab}$ act by commutators.

It can be argued that the contributions from all the terms in the Laplacian $\Omega$ except the first are subleading \cite{CastroVillarreal:2004vh}. In particular, the effect of the quadratic term in (\ref{YM}) (the $\beta-$term) is only to extend slightly the region of the ${\rm AdS}^2$ phase. Thus, the effective potential (per one degree of freedom) in the configuration (\ref{sp0}) is given explicitly by
\begin{eqnarray}
2\frac{V}{N_k^2}=\tilde{\kappa}^4 \bigg[\frac{1}{4}\varphi^4-\frac{1}{3}\varphi^3+\frac{1}{2}\tau\varphi^2\bigg]+\log\varphi^2.\label{sp3}
\end{eqnarray}
The scaled coupling constant $\tilde{\kappa}$ is given explicitly by
\begin{eqnarray}
\tilde{\kappa}^4=4\kappa^4 k(k-1).
\end{eqnarray}
This is the collapsed coupling constant in the large $k$ or equivalently $N_k=2k-1$ limit (commutative limit). Indeed, by expanding the IKKT-type Yang-Mills matrix model (\ref{YM}) around the background (\ref{sp0})  we obtain a $U(1)$ gauge field on the noncommutative anti-de Sitter space ${\rm AdS}^2_{\theta}$ where the gauge coupling constant  $g$ is inversely proportional to $\tilde{\kappa}^2$. The complete gauge dynamics found when expanding around the background (\ref{sp0}) is a noncommutative $U(1)$ gauge field  on ${\rm AdS}^2_{\theta}$ coupled to a scalar field normal to ${\rm AdS}^2_{\theta}$, i.e. a noncommutative Higgs system on ${\rm AdS}^2_{\theta}$.

This is very similar to what happens in the case of the fuzzy sphere $\mathbb{S}^2_N$ where there it is also found that  a UV-IR mixing behavior exists in the large $N$ limit \cite{CastroVillarreal:2004vh}.  However, in the case of noncommutative  anti-de Sitter space ${\rm AdS}^2_{\theta}$ the spectrum of the Laplacian operator on noncommutative ${\rm AdS}^2_{\theta}$ is found to be exactly identical to the spectrum of the Laplacian operator on commutative ${\rm AdS}^2$ and as a consequence it is conjectured that there is no UV-UR mixing on ${\rm AdS}^2_{\theta}$.

In summary, the three results (\ref{sp1}), (\ref{sp2}) and (\ref{sp3}) are exactly identical to those obtained in the fuzzy sphere case  \cite{Delgadillo-Blando:2012skh} and as a consequence we have an exact analogy between the phase diagrams in the two cases. 

The commutative limit of the  above IKKT-type Yang-Mills matrix model (\ref{YM}) with an emphasis on the gravitational degrees of freedom (emergent  matrix or noncommutative gravity) can also be pursued along the lines of \cite{Steinacker:2008ri,Steinacker:2010rh}.

\section{${\rm AdS}_2$ black holes and their evaporation}

It is well known that in two dimensions all negatively curved spacetimes are locally  ${\rm AdS}^2$ and thus stable black hole solutions in two dimensions do not exist in a naive way.  This is similar to the fact that in three dimensions all negavtively curved spacetimes are locally ${\rm AdS}^3$ and thus stable black hole solutions in three dimensions do not  also exist in a naive way. However, black hole solutions in two and three dimensions exist. Indeed, in three dimensions the celebrated  BTZ (Banados, Teitelboim and Zanelli) black hole \cite{Banados:1992wn} is a stable black hole solution which differs from ${\rm AdS}^3$ by global identification and in two dimensions the SS (Spradlin and Strominger) black hole \cite{Strominger:1998yg, Spradlin:1999bn} is also a stable black solution which differs from ${\rm AdS}^2$ by global identification. These black holes are therefore locally identical with the corresponding anti-de Sitter spacetimes and differ from these spacetimes only topologically.

As we will see in the following dilaton gravity in two dimensions provides another way of obtaining stable ${\rm AdS}^2$ black holes which are locally identical to ${\rm AdS}^2$ spacetime but differ from it only globally precisely through the value of the dilaton field.

We start with dilaton gravity theory in four dimensions given by the action \cite{Cadoni:1994uf,Cadoni:1993rn}

\begin{eqnarray}
S=\int d^4x \sqrt{-{\rm det}g^{(4)}} e^{-2\phi}(R^{(4)}-F_{\mu\nu}F^{\mu\nu}).\label{dilaton4}
\end{eqnarray}
A closely related low-energy effective actions  of string theory with similar black holes physics are found in \cite{Garfinkle:1990qj,Giddings:1992kn}).

A spherically symmetric non-singular black hole solution of the equations of motion is given by  \cite{Cadoni:1994uf}

\begin{eqnarray}
F_{ij}=\frac{Q_M}{r^2}\epsilon_{ijk}n_k~,~Q_M=\frac{\sqrt{3}}{2}Q.
\end{eqnarray}

\begin{eqnarray}
ds^2=-(1-\frac{r_+}{r})d\tau^2+\frac{dr^2}{(1-\frac{r_+}{r})(1-\frac{r_-}{r})}+r^2d\Omega_2^2.\label{4dBH}
\end{eqnarray}

\begin{eqnarray}
e^{2(\phi-\phi_0)}=\frac{1}{\sqrt{1-\frac{r_-}{r}}}.
\end{eqnarray}
The inner radius $r_-$ and the outer radius $r_+$ are given in terms of the mass $M$ and the charge $Q_M$ of the black hole by the relations \cite{Cadoni:1994uf}

\begin{eqnarray}
2M=r_+~,~ Q^2=r_+r_-.
\end{eqnarray}
The temperature and the entropy of the black hole are given on the other hand by the relations \cite{Cadoni:1994uf}

\begin{eqnarray}
T=\frac{1}{4\pi r_+}\sqrt{1-\frac{r_-}{r_+}}~,~S_{\rm entro}=\pi r_+^2
\end{eqnarray}
The extremal limit $T\longrightarrow 0$ of this black hole configuration is then given by $r_+=r_-=Q$ or equivalently $M=Q/2$.

The spatial sections of this black hole solution coincide  with those of the Reissner-Nordstrom black hole \cite{RN}. However, this solution corresponds to a non-singular black hole where the spacetime manifold is cut at $r=r_-$ while it is asymptotically flat. Indeed, the maximal extension of this metric yields a Penrose diagram identical to that of the Schwarzschild solution except that the singularity $r=0$ is replaced by the boundary of the manifold at $r=r_-$ \cite{Cadoni:1994uf}.

Furthermore, it is not difficult to show that the near-horizon geometry of this extremal black hole is given by ${\rm AdS}^2\times\mathbb{S}^2$.

By performing spherical reduction of the action (\ref{dilaton4}) on a sphere of constant radius $r=Q$ we obtain the action (with $\Lambda=1/2Q$)

\begin{eqnarray}
S&=&4\pi Q^2 \int d^2x \sqrt{-{\rm det}g^{(2)}} e^{-2\phi}(R^{(2)}+2\Lambda^2).
\end{eqnarray}
This is called the Jackiw-Teitelboim action \cite{JT} which is one of the most important dilatonic gravity models in two dimensions.

The most general solution of the equations of motion stemming from the Jackiw-Teitelboim action   (see \cite{Cadoni:1993rn} and references therein) is given by  the metric field (in the so-called Schwarzschild coordinates) and the dilaton field (with $\Phi=\exp(-2\phi)$)

\begin{eqnarray}
ds^2=-(\Lambda^2r^2-a^2)d\tau^2+\frac{dr^2}{\Lambda^2r^2-a^2}.
\end{eqnarray}

\begin{eqnarray}
e^{2(\phi-\phi_0)}=\frac{1}{\Lambda r}\iff \Phi=e^{-2\phi}=\Phi_0\Lambda r.\label{dilaton}
\end{eqnarray}
We can set $\phi_0=0$ and $\Phi_0=1$ for simplicity. The parameter $a^2$ in the metric is an integration constant related to the mass $M$ of the solution by the relation

\begin{eqnarray}
M=\frac{\Lambda}{2}a^2\Phi_0.\label{temp0}
 \end{eqnarray}
The above metric corresponds, for all values of $a^2$, to a two-dimensional spacetime with a constant negative curvature $R=-2\Lambda^2$, i.e. an anti-de Sitter spacetime ${\rm AdS}^2$. Furthermore, it was shown in \cite{Cadoni:1994uf} that this metric in Schwarzschild coordinates describes the two-dimensional sections of the extremal four-dimensional black hole given in (\ref{4dBH}).

The solution for $a^2=0$ is exactly ${\rm AdS}^2$ spacetime and it plays the role of the ground state of the theory. For example, this solution has mass $M=0$ and the mass of the other solution will be computed with respect to this ground state.

The solution $a^2>0$ is an ${\rm AdS}^2$ black hole with a horizon at $r_H=a/\Lambda$. This spacetime can not be distinguished locally from the pure ${\rm AdS}^2$ solution with $a^2=0$ since there exists a coordinate transformation connecting the two solutions.

However, although the solution with $a^2=0$ (${\rm AdS}^2$ spacetime) is locally equivalent to the solution with $a^2 > 0$ (${\rm AdS}^2$ black hole) these two solutions are globally different due precisely to the behavior of the dilaton field  which effectively sets the boundary condition on the spacetime.

Furthermore, this black hole solution must be cutoff at $r=0$ otherwise the dilaton field $\Phi=\exp(-2\phi)$ will become negative when we maximally extend the corresponding metric beyond $r=0$ which in turn will translate in four dimensions into a negative value for the area of the transverse sphere which is physically unacceptable. Therefore $r=0$ is a boundary for the   ${\rm AdS}^2$ black hole with $a^2>0$ corresponding to the boundary $r=r_-$ of the extremal four-dimensional regular black hole (\ref{4dBH}). The temperature and the entropy of this ${\rm AdS}^2$ black hole can be computed in the usual way and one finds \cite{Cadoni:1994uf}

\begin{eqnarray}
T=\frac{a\Lambda}{2\pi}~,~S_{\rm entro}=4\pi \sqrt{\frac{\Phi_0M}{2\Lambda}}.\label{temp}
\end{eqnarray}
The relationship between the ${\rm AdS}^2$ spacetime corresponding to the solution $a^2=0$ (denoted from now on by ${\rm AdS}_0$) and the ${\rm AdS}^2$ black hole corresponding to the solution $a^2> 0$ (denoted from now on by ${\rm AdS}_+$) is identical to the relationship between the the two-dimensional Minkowski spacetime and the Rindler wedge. Indeed, the parameter $a^2$ in our case (which is proportional to the mass of the black hole)   is the analogue of the acceleration $\alpha$ with which the Rindler observer is uniformly accelerating in Minkowski spacetime creating thus a horizon at $z=|t|$ separating the Rindler wedge from the rest of Minkowski spacetime.





Furthermore, a non-zero value of the parameter $a^2$ is associated with the existence of a horizon at $r=r_H=a/\Lambda$ separating the exterior $r> r_H$ of the black hole ${\rm AdS}_+$ from its interior region $0< r< r_H$. The exterior $r>r_H$ of the black hole ${\rm AdS}_+$ is  described by the light-cone coordinates $\sigma^{\pm}$ given by (with $\sigma$ being the tortoise coordinate $r_*$)

\begin{eqnarray}
\sigma^{\pm}=\tau \pm \sigma~,~\sigma=-\frac{1}{a\Lambda}{\rm arctanh}(\frac{a}{\Lambda r}).\label{dilaton1}
\end{eqnarray}
The black hole coordinates $\tau$ and $\sigma$ are defined in the range $-\infty< \tau < +\infty$ and $0<\sigma< \infty$. The metric on the ${\rm AdS}^2$ black hole ${\rm AdS}_+$ is then given by 

\begin{eqnarray}
ds^2=\frac{a^2}{\sinh^2 a\Lambda\sigma}(-d\tau^2+d\sigma^2).\label{coc0}
\end{eqnarray}
The asymptotic behavior of this ${\rm AdS}^2$ black hole ${\rm AdS}_+$ is given by the ${\rm AdS}^2$ spacetime ${\rm AdS}_0$ which can be fully covered by the light-like coordinates $z^{\pm}=t\pm z$ where $t$ and $z$ are defined by
\begin{eqnarray}
t=\frac{1}{a\Lambda}e^{a\Lambda \tau}\cosh a\Lambda\sigma~,~z=\frac{1}{a\Lambda}e^{a\Lambda \tau}\sinh a\Lambda\sigma.\label{coc1}
\end{eqnarray}
The AdS coordinates $t$ and $z$ are defined in the range $-\infty< t< +\infty$ and $0< z< \infty$. The metric  on ${\rm AdS}_0$ is then of the Poincare form, viz

\begin{eqnarray}
ds^2=\frac{1}{\Lambda^2 z^2}(-dt^2+dz^2).\label{coc2}
\end{eqnarray}
By comparing with (\ref{met2}) we obtain the identification $\Lambda=1/R$. The change of coordinates (\ref{coc1}) which relates the two sets of coordinates $z^{\pm}$ and $\sigma^{\pm}$  (although in our case it does not correspond to any motion of physical observers and therefore is connecting two different manifolds) play exactly the role of the boost which connects the Minkwoski coordinates $(t,x)$ to the Rindler coordinates $(\eta,\sigma)$. This change of coordinates also takes the form
\begin{eqnarray}
z^{\pm}=\frac{1}{a\Lambda}e^{a\Lambda\sigma^{\pm}}.\label{coc3}
\end{eqnarray}
These coordinates define region or quadrant I of ${\rm AdS}_0$ (the exterior of the black hole) in which the timelike Killing vector field (which generates boosts in the $z-$direction) is $\partial_{\tau}$. This Killing vector field is future-directed.  Thus, the Killing horizons lie at $z=\pm t$.

The region or quadrant II of ${\rm AdS}_0$ in which the timelike Killing vector field (given here by $\partial_{-\tau}=-\partial_{\tau}$) is past-directed is given by the coordinates

\begin{eqnarray}
z^{\pm}=-\frac{1}{a\Lambda}e^{a\Lambda\sigma^{\pm}}.\label{coc4}
\end{eqnarray}
The change of coordinates $(t,z)\longrightarrow (\tau,\sigma)$ in this region is given by

\begin{eqnarray}
t=-\frac{1}{a\Lambda}e^{a\Lambda \tau}\cosh a\Lambda\sigma~,~z=-\frac{1}{a\Lambda}e^{a\Lambda \tau}\sinh a\Lambda\sigma.\label{coc5}
\end{eqnarray}
Hence ${\rm AdS}_+$ covers the region of ${\rm AdS}_0$ given by the union of the two quadrants I and II which is specified by the condition $z^+z^-\geq 0$ with the black horizon defined by the condition $z^+z^-=0$.

Therefore, quantization of fields and semi-classical considerations of Hawking radiation in the ${\rm AdS}_+$ background with ${\rm AdS}_0$ taken as a ground state should then proceed along the same steps as the analogous calculation performed in the Rindler wedge with the Minkowski spacetime taken as a ground state.

However, we should also recall that anti-de Sitter spacetime is not a globally hyperbolic space and thus one should be careful with the boundary conditions at infinity (transparent boundary conditions) as discussed for example in \cite{Cadoni:1993rn}. See also the exciting recent work on the black hole information loss problem based on the Almheiri-Polchinski (AP) model \cite{Almheiri:2014cka}.

Thus, there will be two vacuum states $|0_+\rangle$ and $|0_0\rangle$  corresponding to the two different spacetimes (observers) ${\rm AdS}_+$ and  ${\rm AdS}_0$ and as a consequence a thermal radiation will be observed in each vacuum state by the other observer associated with the other spacetime which is precisely Hawking radiation. The corresponding annihilation operators are defined by 

\begin{eqnarray}
\hat{a}_k|0_0\rangle=0.
\end{eqnarray}

\begin{eqnarray}
\hat{b}_k^{(1)}|0_+\rangle=\hat{b}_k^{(2)}|0_+\rangle=0.
\end{eqnarray}
The ${\rm AdS}_+$ number operator in region I is then defined by

\begin{eqnarray}
\hat{N}_R^{(1)}(k)=\hat{b}_k^{(1)+}\hat{b}_k^{(1)}.
\end{eqnarray}
We can now immediately compute the expectation value of this number operator in region I in the anti-de Sitter vacuum $|0_0\rangle$ to find

\begin{eqnarray}
\langle 0_0|\hat{N}_R^{(1)}(k)|0_0\rangle
&=&\frac{1}{\exp(\frac{2\pi\omega}{a\Lambda})-1}\delta(0).
\end{eqnarray}
This is a blackbody Planck spectrum corresponding to the temperature

\begin{eqnarray}
T=\frac{a\Lambda}{2\pi}.
\end{eqnarray}
The Hawking effect \cite{Hawking:1974sw} computed in this case is mathematically equivalent to the Unruh effect \cite{Unruh:1976db}.

In the first appendix  a much more detailed discussion of the Hawking process in ${\rm AdS}^2$ spacetime is included.

\section{Noncommutative black holes and matrix models}
We go to Euclidean signature by the usual Wick rotation $\tau\longrightarrow -i\tau$.

By employing the definition of the dilaton field solution (\ref{dilaton}), the definition of the tortoise coordinate $\sigma$ in terms of the radial coordinate $r$ given by equation (\ref{dilaton1}), the change of coordinates (\ref{coc1}), the definition of the global coordinates $X_2$ in terms of the Poincare coordinates $t$ and $z$ given by the second equation of (\ref{global}), the fact that $\Lambda=1/R$, and the definition of the ${\rm AdS}^2$ black hole temperature  $T$ given by equation (\ref{temp}) we obtain
\begin{eqnarray}
\Phi=2\pi T X_2.
\end{eqnarray}
This is a simple relation giving the dilaton field of the ${\rm AdS}^2$ black hole in terms of the global coordinate $X_2$ of the ${\rm AdS}^2$ spacetime which will be our starting point to construct noncommutative ${\rm AdS}^2_{\theta}$ black hole (noncommutative geometry in the presence of black hole configurations) and their quantum gravity analogue (matrix model defining quantization about noncommutative black hole backgrounds).

As we have seen, after canonical quantization the global coordinates $X_a$ become operators $\hat{X}_a$ in a Hilbert space satisfying the relations
 (\ref{45}), (\ref{46}) and (\ref{47}). The dilaton field $\Phi$ becomes therefore an operator $\hat{\Phi}$ in this same Hilbert space given by the relation
 \begin{eqnarray}
\hat{\Phi}=2\pi T \hat{X}_2.\label{48}
\end{eqnarray}
The noncommutative ${\rm AdS}^2_{\theta}$ black hole is given by the four matrices $\hat{X}^a$, $\Phi$ satisfying the constraints  (\ref{45}), (\ref{46}), (\ref{47}) and (\ref{48}).

Recall the metric $\eta_{\mu\nu}=(-1,+1,+1)$ and the Levi-Civita tensor is defined by $\epsilon^{123}=+1$. We can then write the above equation as the following three commutators (with $\kappa_1=2\pi T \kappa$)
\begin{eqnarray}
[\hat{\Phi},\hat{X}^1]=-i \kappa_1 \hat{X}^3.
\end{eqnarray}
\begin{eqnarray}
[\hat{\Phi},\hat{X}^2]=0.
\end{eqnarray}
\begin{eqnarray}
[\hat{\Phi},\hat{X}^3]=-i \kappa_1 \hat{X}^1.
\end{eqnarray}
We think of these three equations as the equations of motion for the matrix dilaton field $\hat{\Phi}$. A supplementary matrix model which yields (\ref{48}) as a global minimum without altering the solution given by (\ref{45}), (\ref{46}) and (\ref{47}) is provided by
\begin{eqnarray}
S_{\Phi}&=&\frac{N}{2} {\rm Tr}\bigg[\big(i[\hat{\Phi},D^1]-\kappa_1 D^3\big)\big(i[\hat{\Phi},D_1]+\kappa_1 D_3\big)\nonumber\\
&+&\big(i[\hat{\Phi},D^2]\big)\big(i[\hat{\Phi},D_2]\big)\nonumber\\
&+&\big(i[\hat{\Phi},D^3]-\kappa_1 D^1\big)\big(i[\hat{\Phi},D_3]+\kappa_1 D_1\big)\bigg]\nonumber\\
&=&\frac{N}{2} {\rm Tr}\bigg[-[\hat{\Phi},D^a][\hat{\Phi},D_a]+4i\kappa_1\hat{\Phi}[D^1,D^3]\nonumber\\
&-&\kappa_1^2D^1D_1-\kappa_1^2D^3D_3\bigg].\label{YM2}
\end{eqnarray}
The full matrix action describing quantum gravitational fluctuations about the noncommutative ${\rm AdS}^2_{\theta}$ black hole is given by the sum of the two actions (\ref{YM}) and (\ref{YM2}).  By defining $D_4=D^4=\Phi$ this complete matrix action is seen to be a $4-$dimensional Yang-Mills matrix model with mass deformations which are invariant only under the subgroup of $SO(1,2)$ pseudo-rotations around $X_2$. We get the action
\begin{eqnarray}
S[D]&=&N {\rm Tr}\bigg(-\frac{1}{4}[D_{\mu},D_{\nu}][D^{\mu},D^{\nu}]+2i\kappa D^2[D^1,D^3]\nonumber\\
&+&2i\kappa_1\hat{\Phi}[D^1,D^3]+\beta D_aD^a\nonumber\\
&-&\frac{1}{2}\kappa_1^2D_1D^1-\frac{1}{2}\kappa_1^2D_3D^3\bigg).\label{mp}
\end{eqnarray}
We are interested in the phase structure of the model with the parameters
\begin{eqnarray}
\beta=0~,~\kappa\neq 0~,~\kappa_1=2\pi T\kappa.
\end{eqnarray}
We have then two parameters $\kappa$ and $\kappa_1$ or equivalently $\kappa$ and $T$. We will fix the value of the parameter $\kappa_1$ so that an increase in the actual temperature $T$ of the black hole corresponds to a decrease in the inverse gauge coupling constant $\kappa$ \cite{Delgadillo-Blando:2007mqd}.   

Yang-Mills matrix models are generally characterized by two phases: 1) A Yang-Mills phase and 2) A geometric phase due to the Myers terms, i.e. the cubic terms in (\ref{mp}). See \cite{Ydri:2020fry} and references therein.

The first cubic term in (\ref{mp}) is responsible for the condensation of the geometry (${\rm AdS}^2_{\theta}$ space) whereas the second cubic term in (\ref{mp}) is responsible for the stability of the noncommutative black hole solution in the ${\rm AdS}^2_{\theta}$ space.  There should then be two critical points ${\kappa}_{1 \rm cr}$ and  ${\kappa}_{2 \rm cr}$ corresponding to the vanishing of the tow cubic expectation values $\langle i {\rm Tr}  D^2[D^1,D^3]\rangle$ and $\langle i {\rm Tr} \Phi [D^1,D^3]\rangle$ respectively. By assuming now that $\kappa_1>\kappa$ we have ${\kappa}_{2 \rm cr}> {\kappa}_{1 \rm cr}$ and we have the following three phases (a gravitational phase, a geometric phase and a Yang-Mills phase):
\begin{itemize}
\item For $\kappa> \kappa_{2 \rm cr}$ we have a black hole phase where $\hat{\Phi}\propto \hat{X}_2$. This is the gravitational phase in which we have a stable noncommutative ${\rm AdS}^2_{\theta}$ black hole.
\item For ${\kappa}_{1 \rm}< \kappa < {\kappa}_{2 \rm}$ the second cubic term effectively vanishes and condensation is now only driven by the first cubic term. In other words, although we have in this phase $\hat{\Phi}\sim 0$ the three matrices $D_a$ still fluctuate about the noncommutative  ${\rm AdS}^2_{\theta}$ background given by the matrices  $\hat{X}_a$. This is the geometric phase.
\item As we keep decreasing the temperature, i.e. increasing $\kappa$ below $\kappa_1$ the   ${\rm AdS}^2_{\theta}$ background evaporates in a transition to a pure Yang-Mills matrix model. In this Yang-Mills phase, all four matrices become centered around zero. 
\end{itemize}
This is the Hawking process as it presents itself in the Yang-Mills matrix model, i.e. as an exotic line of discontinuous transitions with a jump in the entropy, characteristic of a 1st order transition, yet with divergent critical fluctuations and a divergent specific heat characteristic of a 2nd order transition.

\section{Dilaton gravity in two dimensions as a non-linear sigma model}

An alternative way of recasting the Hawking process as a phase transition starts from  the Jackiw-Teitelboim action which can be rewritten as a non-linear sigma model. The vacuum expectation value  of the dilaton field should vanish in the pure ${\rm AdS}^2$ background while it will take a non-trivial value in the  ${\rm AdS}^2$ black hole background. In this section we will rewrite the Jackiw-Teitelboim action as a non-linear sigma model, then write down its noncommutative analogue and briefly discuss the corresponding phase structure.

The most general dilaton  gravity action  in two dimensions is given by (after an appropriate Weyl rescaling of the metric)

\begin{eqnarray}
S=\int d^2x\sqrt{-{\rm det} g}(\Phi R+V(\Phi)).\label{dg}
\end{eqnarray}
For example, the Jackiw-Teitelboim model coressponds to the potential $V(\Phi)=2\Lambda^2\Phi$.

There are two  physical degrees of freedom in this theory: one given by the dilaton field $\Phi$ and one contained in the metric $g_{\mu\nu}$. Indeed, in two dimensions the metric is of the generic form

\begin{eqnarray}
g_{\mu\nu}=\rho\left( \begin{array}{cc}
\alpha^2-\beta^2 &\beta \\
 \beta & -1 \end{array}\right).
\end{eqnarray}
The "lapse function" $\alpha=\alpha(t,x)$ and the "shift vector" $\beta=\beta(t,x)$ are non-dynamical variables here whereas the scale factor $\rho=\rho(t,x)$ is the only metric dynamical variable in two dimensions.

The metric in two dimensions can also be rewritten in the form (with $u=(t+x)/2$ and $v=(t-x)/2$ being the conformal light-cone coordinates)

\begin{eqnarray}
ds^2=4\rho(u,v)dudv.\label{myu}
\end{eqnarray}
In other words, the metric (any metric in two dimensions) is locally conformally flat.

The above model (\ref{dg}) is completely integrable which means that it can be completely solved in terms of free fields \cite{Cavaglia:1998xj}.

We start from the equations of motion

\begin{eqnarray}
(\nabla_{(\mu}\nabla_{\nu)}-g_{\mu\nu}\nabla_{\sigma}\nabla^{\sigma})\Phi+\frac{1}{2}g_{\mu\nu}V=0. \label{em1}
\end{eqnarray}

 \begin{eqnarray}
R+\frac{dV}{d\Phi}=0.\label{em2}
\end{eqnarray}
Define also

\begin{eqnarray}
H(g,\Phi)= \nabla_{\sigma}\Phi\nabla^{\sigma}\Phi\ne 0
\end{eqnarray}
If $H\ne 0$ then the equation of motion (\ref{em2}) is automatically satisfied if the equation of motion (\ref{em1}) is satisfied \cite{Cavaglia:1998xj}.

The so-called  Bäcklund transformation \cite{filippov} allows us to transform the interacting fields $\Phi$ and $g_{\mu\nu}$ into the free fields $M$ and $\psi$ as follows

\begin{eqnarray}
\nabla_{\mu}\psi=\frac{\nabla_{\mu}\Phi}{H(g,\Phi)}.
\end{eqnarray}

\begin{eqnarray}
M=N(\Phi)-H(g,\Phi)~,~N(\Phi)=\int^{\Phi}d\Phi^{\prime} V(\Phi^{\prime}).\label{ADM}
\end{eqnarray}
Indeed, $\psi$ and $M$ are free fields since $\nabla_{\mu}\nabla^{\mu}\psi=0$ and $\nabla_{\mu}M=0$. These equations of motion are equivalent to the original equations of motion (\ref{em1}) and (\ref{em2}). In particular, the second equation $\nabla_{\mu}M=0$ means that the field $M$ is actually a locally conserved quantity (the ADM mass). In the conformal light-cone coordinates the solution of these free equations of motion is trivially given by

\begin{eqnarray}
\psi=F(u)+G(v)~,~M=M_{\rm ADM}.
\end{eqnarray}
We can explicitly determine the dependence of the original fields $\Phi$ and $\rho$ on the free fields $\psi$ and $M$ to be given by \cite{Cavaglia:1998xj}

\begin{eqnarray}
\frac{d\psi}{d\Phi}=\frac{1}{N(\Phi)-M}~,~\rho=(N(\Phi)-M)\partial_u\psi\partial_v\psi.
\end{eqnarray}
The metric (\ref{myu}) becomes then

\begin{eqnarray}
ds^2=4(N(\Phi)-M)dFdG.
\end{eqnarray}
In other words, $F$ and $G$ appear as conformal light-cone coordinates. Thus, we introduce together with $\psi=F+G$ a timelike coordinate $T$ by $T=F-G$. The metric becomes

\begin{eqnarray}
ds^2=-(N(\Phi)-M)dT^2+\frac{d\Phi^2}{(N(\Phi)-M)}.\label{mele}
\end{eqnarray}
The dilaton field $\Phi$ appears therefore as a radial coordinate and it is the only dynamical variable appearing in the above general solution. In some sense this result generalizes Birkhoff theorem (in Einstein gravity in four dimensions with spherical symmetry the only local constant of the motion is the Schwarzschild mass).

As an example, we consider the dimensional reduction of Einstein gravity in four dimensions on a sphere of radius $R^2=4\Phi$ (spherical reduction is consistent in the case of maximal rotational invariance) . The resulting action takes the form (\ref{dg}) with a potential $V(\Phi)=1/2\sqrt{\Phi}$ and as a consequence the metric element (\ref{mele}) reduces to the radial part of the Schwarzschild solution.

By taking the derivative of equation (\ref{ADM}) we obtain

\begin{eqnarray}
\nabla_{\mu}M=V.\nabla_{\mu}\Phi-\nabla_{\mu}H.
\end{eqnarray}
By employing this result we can express the potential $V$ in terms of $M$ and $\nabla_{\mu}\Phi$ as follows

\begin{eqnarray}
V=\frac{1}{H}\nabla^{\mu}\Phi\nabla_{\mu}M+\frac{1}{H}\nabla^{\mu}\Phi\nabla_{\mu}H.\label{piece1}
\end{eqnarray}

From the other hand, the Ricci scalar in two dimensions is locally given by a total divergence, viz

\begin{eqnarray}
R=2\nabla_{\mu}A^{\mu}~,~A^{\mu}=\frac{\nabla^{\mu}\nabla^{\nu}\chi.\nabla_{\nu}\chi-\nabla_{\nu}\nabla^{\nu}\chi.\nabla^{\mu}\chi}{\nabla^{\rho}\chi.\nabla_{\rho}\chi}.
\end{eqnarray}
In this equation $\chi$ is an arbitrary scalar field which we choose to be $\chi=\Phi$. We then compute

\begin{eqnarray}
 \Phi.R&=&2\Phi.\nabla_{\mu}A^{\mu}\nonumber\\ &=&2\nabla_{\mu}(\Phi.A^{\mu})-2\nabla_{\mu}\Phi.A^{\mu}\nonumber\\
&=& 2\nabla_{\mu}(\Phi.A^{\mu})-\frac{2}{H}\nabla_{\mu}\Phi.(\frac{1}{2}\nabla^{\mu}H-\nabla_{\nu}\nabla^{\nu}\Phi.\nabla^{\mu}\Phi)\nonumber\\
&=&2\nabla_{\mu}(\Phi A^{\mu}+\nabla^{\mu}\Phi)-\frac{1}{H}\nabla_{\mu}\Phi.\nabla^{\mu}H.\label{piece2}
 \end{eqnarray}
By putting (\ref{piece1}) and (\ref{piece2}) together  and using the fact that $H=N(\Phi)-M$ we can show that the original dilaton gravity action (\ref{dg}) takes the form (up to a surface term)

\begin{eqnarray}
S=\int d^2x\sqrt{-{\rm det} g}\frac{\nabla_{\mu}M\nabla^{\mu}\Phi}{N(\Phi)-M}.
\end{eqnarray}
This is a non-linear sigma model. The mass functional $M$ is constant on the classical orbit equal to the ADM mass of the black hole given by (\ref{temp}), viz \cite{Cadoni:2000wa}

\begin{eqnarray}
M=\frac{2\pi^2}{\Lambda}T^2.
\end{eqnarray}
We switch now to Euclidean signature. Furthermore, we will implement the constraint (\ref{ADM}) through a Gaussian term in the path integral. The action reads explicitly

\begin{eqnarray}
S&=&\int d^2x\sqrt{{\rm det} g}\bigg[\frac{\partial_{\mu}M\partial^{\mu}\Phi}{N(\Phi)-M}\nonumber\\
&+&m^2\bigg(M-N(\Phi)+\partial_{\mu}\Phi\partial^{\mu}\Phi\bigg)^2\bigg].
\end{eqnarray}
In the limit $m^2\longrightarrow\infty$ we recover the  constraint (\ref{ADM}). We can write down the noncommutative analogue of this action in a straightforward way. However, we are more interested in the phase structure of this model corresponding to a constant dilaton configuration, i.e. $\Phi={\rm constant}$. In this case we are really dealing with a potential model given explicitly by 
\begin{eqnarray}
V&=&\int d^2x\sqrt{{\rm det} g}\bigg[m^2\bigg(M-N(\Phi)\bigg)^2\bigg].
\end{eqnarray}
The noncommutative analogue of this potential is simply given by 
\begin{eqnarray}
V&=&2\pi R\kappa {\rm Tr}\bigg[m^2\bigg(\hat{M}-N(\hat{\Phi})\bigg)^2\bigg].
\end{eqnarray}
For the case of the Jackiw-Teitelboim action we have $V=2\Lambda^2\hat{\Phi}$ and $N(\hat{\Phi})=\Lambda^2\hat{\Phi}^2$. This is (after regularization) a real quartic matrix model of the form \cite{Brezin:1977sv,Shimamune:1981qf}
\begin{eqnarray}
V=N_k{\rm Tr}(\mu\hat{\Phi}^2+g\hat{\Phi}^4).
\end{eqnarray}
The parameters $\mu$ and $g$ are given explicitly by
\begin{eqnarray}
\mu=-2\pi M.(\frac{m^2}{k^2})~,~g=\pi\Lambda^2.(\frac{m^2}{k^2}).
\end{eqnarray}
The one-cut solution (disordered phase) in which the dilaton field $\hat{\Phi}$ fluctuates around zero, i.e. with eigenvalues supported in a single interval  occurs for all values of the mass $M$ satisfying

\begin{eqnarray}
M\leq M_*~,~M_*=\frac{\Lambda}{\sqrt{\pi}}\frac{k}{m}.
\end{eqnarray}
The large mass limit $m^2\longrightarrow \infty$ is then required to be correlated with the commutative limit $k\longrightarrow \infty$. This critical value $M_*$ is consistent with the black hole mass (\ref{temp0}). In other words, the dilaton  field is zero when the mass parameter $M$ of the matrix model is lower than the black hole mass given by (\ref{temp0}), i.e. the black hole has already evaporated and we have only pure ${\rm AdS}^2_{\theta}$ space in this phase.

Above the line $M=M_*$ we have a two-cut solution (non-uniform ordered phase) where the dilaton field $\hat{\Phi}$ is proportional to the idempotent matrix $\gamma$ (which satisfies $\gamma^2={\bf 1}$).  In this two-cut phase we have therefore a non-trivial dilaton field corresponding to a stable ${\rm AdS}^2_{\theta}$  black hole. The transition at $M=M_*$ between the the one-cut and two-cut phases is third order.

\section{Star products in black hole configurations}
The Moyal-Weyl star product and the Weyl map were always very powerful tools for semi-classical computations and for the intuitive visualization of noncommutative geometry.

Another crucial tool is Darboux theorem which states that for every Poisson manifold with an invertible Poisson structure $\theta^{ij}$, i.e. for every symplectic manifold  there exists a coordinate system $\bar{x}$ in which the noncommutativity parameter $\bar{\theta}^{ij}$ is constant. Thus, according to Darboux theorem every noncommutative or quantized symplectic manifold is locally a noncommutative Moyal-Weyl space. This is in fact the equivalence principle for symplectic geometry and its corresponding emergent noncommutative gravity as discussed in \cite{Lee:2010zf1}. See also \cite{Blaschke:2010ye} for the use of Darboux theorem in determining the symplectic structure (noncommutativity parameter) in matrix models.

In this section we will employ Darboux theorem to write down the Moyal-Weyl star product for pure noncommutative ${\rm AdS}^2_{\theta}$ and then we will write down the Moyal-Weyl product for the   noncommutative ${\rm AdS}^2_{\theta}$ black hole by employing one more time Darboux theorem, i.e. by employing the fact that ${\rm AdS}^2$ spacetime and the corresponding ${\rm AdS}^2$ black hole are related by a coordinate transformation. Indeed, the dilaton field enters only globally/topologically and thus it does not affect the star product which is a purely local object.

Let us recall the metric and the fundamental Poisson bracket in the  Poincare coordinates $(t,z)$ given respectively by
\begin{eqnarray}
     ds^2=\frac{R^2}{z^2}(dz^2+dt^2).\label{met}
    \end{eqnarray}
    We compute immediately the Poisson bracket
     \begin{eqnarray}
       &&\{t,z\}=-\frac{R^3}{(X_1-X_3)^3}\{X_2,X_1-X_3\}=\frac{\kappa}{R}z^2.\label{Poi}\nonumber\\
     \end{eqnarray}
 Formally this Poisson bracket (\ref{Poi}) can be derived from the symplectic structure (with $x^1=t$ and $x^2=z$)
    \begin{eqnarray}
     \omega=\frac{1}{2}(\theta^{-1})_{ij}dx^i\wedge dx^j\equiv -\frac{1}{\kappa}\frac{R}{z^2}dt\wedge dz.
    \end{eqnarray}
    The  corresponding Poisson bracket is given by
    \begin{eqnarray}
      \{f,g\} &=&\theta^{ij}\partial_if\partial_jg\nonumber\\
      &=&\kappa\frac{z^2}{R}(\partial_tf\partial_z g-\partial_zf\partial_tg).
    \end{eqnarray}
    Thus, the Poisson structure or noncommutativity parameter $\theta$ is an $x-$dependent tensor given explicitly by
    \begin{eqnarray}
      \theta^{ij}=\frac{\kappa}{R}z^2\epsilon^{ij}.
      \end{eqnarray}
    It is well known from deformation quantization \cite{Kontsevich:1997vbv2} that these Poisson brackets define the leading correction in the formal expansion of the star product. Explicitly, we should have (recall that upon quantization $i\{f,g\}$ becomes the commutator $[\hat{f},\hat{g}]$)
     \begin{eqnarray}
       f*g(x)&=&f(x)g(x)+\frac{i}{2}\{f,g\}+O(\theta^2)\nonumber\\
       &=&f(x)g(x)+\frac{i}{2}\theta^{ij}\partial_if\partial_jg+O(\theta^2).
     \end{eqnarray}
     We can also compute the Poisson bracket (with $u=1/z$)
 \begin{eqnarray}
      &&\{t,u\}=\frac{1}{R(X_1-X_3)}\{X_2,X_1-X_3\}=-\frac{\kappa}{R}.\nonumber\\
    \end{eqnarray}
    The parameter $\kappa/R$ is dimensionless and thus $z$ is the radial coordinate of anti-de Sitter spacetime  ${\rm AdS}^2$ whereas $u$ is the energy scale of the conformal field theory   ${\rm CFT}_1$ living on its boundary located at $z\longrightarrow 0$. We observe from this second Poisson bracket that $t$ and $u$ are conjugate variables but they are not necessarily canonical variables because their phase phase is not the full $\mathbb{R}^2$ but it is $\mathbb{R}\times\mathbb{R}_+$ \cite{Kastrup:2003fs}. A set $(x,y)$ of canonical coordinates on  ${\rm AdS}^2$ with phase space given by $\mathbb{R}^2$  is given by
    \begin{eqnarray}
     z=R\exp(-x)~,~t=\exp(-x)y.
    \end{eqnarray}
    We compute immediately the metric
    \begin{eqnarray}
     ds^2=R^2dx^2+(dy-ydx)^2.\label{met1}
    \end{eqnarray}
    We can check quite easily that the fundamental Poisson bracket is given in this case by
    \begin{eqnarray}
      \{x,y\}=\kappa.\label{Poi1}
    \end{eqnarray}
In our case  a coordinate system in which the noncommutative ${\rm AdS}^2_{\theta}$ appears to be a noncommutative Moyal-Weyl space is then already at hand. Indeed, the fundamental Poisson bracket (\ref{Poi1}) in the canonical coordinates $(x,y)$ with metric  (\ref{met1}) can be derived from the symplectic structure (with $\bar{x}^1=x$ and $\bar{x}^2=y$)
    \begin{eqnarray}
     \omega=\frac{1}{2}(\bar{\theta}^{-1})_{ij}d\bar{x}^i\wedge d\bar{x}^j\equiv -\frac{1}{\kappa}dx\wedge dy.
    \end{eqnarray}
    The  corresponding Poisson bracket is given by
    \begin{eqnarray}
      \{f,g\} &=&\bar{\theta}^{ij}\bar{\partial}_if\bar{\partial}_jg=\kappa (\partial_xf\partial_y g-\partial_yf\partial_xg)~,~\bar{\theta}^{ij}=\kappa \epsilon^{ij}.\nonumber\\
    \end{eqnarray}
    And the star product is given by
     \begin{eqnarray}
       f\bar{*}g(\bar{x})&=&f(\bar{x})g(\bar{x})+\frac{i}{2}\bar{\theta}^{ij}\bar{\partial}_if\bar{\partial}_jg\nonumber\\
       &-&\frac{1}{8}\bar{\theta}^{ij}\bar{\theta}^{kl}\bar{\partial}_i\bar{\partial}_kf\bar{\partial}_j\bar{\partial}_lg +O(\bar{\theta}^3).
     \end{eqnarray}
  The canonical coordinates $(x,y)$ play the role of Darboux coordinates on the noncommutative ${\rm AdS}^2_{\theta}$ since $\bar{\theta} $ is a constant tensor \footnote{These Darboux coordinates cover the whole of ${\rm AdS}^2$. The coordinates $(t,u)$ are also Darboux coordinates for which the noncommutativity is constant given by $\tilde{\theta}^{ij}=-\kappa \epsilon^{ij}/R$ but these coordinates cover only half of  ${\rm AdS}^2$.}.  But this star product, again since $\bar{\theta}$ is constant, is precisely given by the Groenewold-Moyal-Weyl star product  \cite{Moyal:1949skv2,Groenewold:1946kpv2} defined by

\begin{eqnarray}
  f\bar{*}g(\bar{x})&=&\exp(\frac{i}{2}\bar{\theta}^{ij}\frac{\partial}{\partial\xi^i}\frac{\partial}{\partial\eta^j})f(\bar{x}+\xi)g(\bar{x}+\eta)|_{\xi=\eta=0}\nonumber\\
  &=&f(\bar{x})\exp\big(\frac{i}{2}\overleftarrow{\bar{\partial}}_i\bar{\theta}^{ij}\overrightarrow{\bar{\partial}}_j\big)g(\bar{x}).
\end{eqnarray}
Under the coordinate transformation $x\longrightarrow \bar{x}$ we have 
\begin{eqnarray}
  \bar{\theta}^{ij}=\theta^{kl}(x)\frac{\partial \bar{x}^i}{\partial x^k}\frac{\partial \bar{x}^j}{\partial x^l}.\label{formula}
\end{eqnarray}
After substitution we get
\begin{eqnarray}
  f\bar{*}g(\bar{x})
  &=&f(x)\exp\big(\frac{i}{2}\overleftarrow{\partial}_m \theta^{mn}(x)\overrightarrow{\partial}_n\big)g(x)\nonumber\\
  &=&fg+\frac{i}{2}\theta^{mn}\partial_mf\partial_n g\nonumber\\
  &-&\frac{1}{8}\partial_p(\theta^{mn}\partial_mf)\partial_n(\theta^{pq}\partial_qg)+O(\theta^3).\label{star2}
\end{eqnarray}
On a quantized Poisson manifold  such as noncommutative ${\rm AdS}^2_{\theta}$ the Poisson structure $\theta$ is generally not a constant tensor and hence the star product is given by   a formal power series in $\theta^{mn}(x)$. The star product which is associative upto the second order in $\theta^{mn}(x)$  is given explicitly by \cite{Kontsevich:1997vbv2}

\begin{eqnarray}
  f*g&=&fg+\frac{i}{2}\theta^{mn}\partial_mf\partial_ng-\frac{1}{8}\theta^{mn}\theta^{pq}\partial_m\partial_pf\partial_n\partial_qg\nonumber\\
  &-&\frac{1}{12}\theta^{mn}\partial_n\theta^{pq}\big(\partial_m\partial_pf\partial_qg+\partial_m\partial_pg\partial_qf\big)+O(\theta^3).\label{star1}\nonumber\\
\end{eqnarray}
The two star products (\ref{star2}) and (\ref{star1}) should be related by a gauge transformation (in the sense of deformation quantization) corresponding to the  coordinate transformation $x\longrightarrow \bar{x}$ \cite{Kontsevich:1997vbv2}. See also the second appendix.

In summary, the formula (\ref{formula}) allows us to compute the symplectic structure (noncommutativity parameter) and as a consequence it allows us to compute the star product of pure noncommutative ${\rm AdS}^2_{\theta}$ in the Poincare coordinates $(t,z)$.

Similar considerations hold in the presence of non-trivial black hole configurations. The metric of the corresponding ${\rm AdS}^2$ black hole is given by
\begin{eqnarray}
ds^2=-(\Lambda^2r^2-a^2)d\tau^2+\frac{dr^2}{\Lambda^2r^2-a^2}.
\end{eqnarray}
The map to a pure  ${\rm AdS}^2$ spacetime is given by ($\sigma$ is the tortoise coordinate)

\begin{eqnarray}
t=-\frac{i}{a\Lambda}e^{-ia\Lambda \tau}\cosh a\Lambda\sigma~,~z=\frac{1}{a\Lambda}e^{-ia\Lambda \tau}\sinh a\Lambda\sigma.\nonumber\\
\end{eqnarray}
The metric becomes then given by (\ref{met}) with $\Lambda=1/R$. The symplectic structure  of pure noncommutative ${\rm AdS}^2_{\theta}$  in the Poincare coordinates $(t,z)$  will then transform under this change of coordinates to the  symplectic structure  of noncommutative ${\rm AdS}^2_{\theta}$ black hole in an obvious way, i.e. as a tensor.

\section{Conclusion}

This article contains the second part of our study in which we attempt a coherent unification between the principles of noncommutative geometry (and their matrix models) from the one hand and the principles of the ${\rm AdS}^{2}/{\rm CFT}_1$ correspondence from the other hand. In this part the main focus has been on constructing noncommutative ${\rm AdS}^2_{\theta}$ black holes and their matrix models.

Noncommutative geometry is understood here as "first quantization" of geometry whereas the corresponding Yang-Mills matrix models provide "quantum gravity" or "second quantization" of the corresponding geometry. The symmetry structure given here by the Lorentz group $SO(1,2)$ is the starting point of the noncommutative geometry but it also  the unifying structure underlying: 1) the ${\rm AdS}^2$ spacetime, 2) the noncommutative ${\rm AdS}^2_{\theta}$ space and 3) the ${\rm CFT}_1$ theory on the boundary given by the dAFF conformal quantum mechanics.

The commutative ${\rm AdS}^2$ black holes are constructed in the two-dimensional Jackiw-Teitelboim dilaton gravity and the corresponding Hawking process is formulated as an Unruh effect between  ${\rm AdS}_0$ (ground state) and ${\rm AdS}_+$ (excited state). The dilaton field plays a fundamental role here in distinguishing between pure ${\rm AdS}^2$ spacetime (${\rm AdS}_0$) where it takes zero value and ${\rm AdS}^2$ black hole spacetime (${\rm AdS}_+$) where it takes a non-trivial value. These two solutions are locally equivalent and they are only globally/topologically differentiated by the value of the dilaton field.

More importantly  we have presented in this article the construction of noncommutative ${\rm AdS}^2_{\theta}$ black hole and its four-dimensional Yang-Mills IKKT-type matrix model which includes two competing Myers term one responsible for the condensation of pure ${\rm AdS}^2_{\theta}$ and the other one responsible for the condensation of the dilaton field. It is argued that the phase diagram of this matrix model features three phases: 1) A
gravitational phase (${\rm AdS}^2_{\theta}$ black hole), 2) A geometric phase (${\rm AdS}^2_{\theta}$ background) and 3) A Yang-Mills phase (no discernible gravitational/geometrical content in the sense described here).

The Hawking process within the Yang-Mills matrix model is therefore given by an exotic line of discontinuous transitions between the gravitational and geometrical phases.

An alternative proposal for noncommutative dilaton gravity is also given in terms of a non-linear sigma model. The vacuum expectation value of the dilaton field should vanish in the pure ${\rm AdS}^2$ background while it will take a non-trivial value in the ${\rm AdS}^2$ black hole background which corresponds in the noncommutative/matrix model to a third order phase transition from a disordered one-cut solution (pure ${\rm AdS}^2_{\theta}$) to a uniform-ordered two-cut solution (${\rm AdS}^2_{\theta}$ black hole).

\section{Hawking process}

The  equations of motion of the Jackiw-Teitelboim action admits the following three solutions:

\begin{itemize}
\item ${\rm AdS}_0$: The solution for $a^2=0$ is pure ${\rm AdS}^2$ spacetime and it plays the role of the ground state of the theory (analogous to Minkowski spacetime). For example, this solution has mass $M=0$ whereas the mass of the other solutions is computed with respect to this ground state.

\item ${\rm AdS}_+$: The solution $a^2>0$ is an ${\rm AdS}^2$ black hole with a horizon at $r_H=a/\Lambda$ and it  can not be distinguished locally from pure ${\rm AdS}^2$ spacetime with $a^2=0$. Indeed, by means of an appropriate change of coordinates we can bring the solution $a^2>0$ into the form of the solution $a^2=0$. The difference between the two cases is strictly topological in character originating from the global properties of the solution encoded in the behavior of the dilaton field.

To see this crucial point more explicitly we consider the coordinate transformation

\begin{eqnarray}
r^{\prime}=a\Lambda \tau r~,~2a\Lambda \tau^{\prime}=\ln\big(\Lambda^2\tau^2-\frac{1}{\Lambda^2 r^2}\big).
\end{eqnarray}
We can then check immediately that

\begin{eqnarray}
ds^2&=-&(\Lambda^2r^{\prime 2}-a^2)d\tau^{\prime 2}+\frac{dr^{\prime 2}}{\Lambda^2r^{\prime 2}-a^2}\nonumber\\
&=&-\Lambda^2r^2 d\tau^2+\frac{dr^2}{\Lambda^2r^2}.
\end{eqnarray}
However, the dilaton field changes in a non-trivial way under the above coordinate transformation, viz

\begin{eqnarray}
\Phi_0\sqrt{\frac{\Lambda^2 r^{\prime 2}}{a^2}-1}e^{-a\Lambda \tau^{\prime}}=\Phi_0\Lambda r.
\end{eqnarray}
Thus, although the solution with $a^2=0$ (pure ${\rm AdS}^2$ spacetime) is locally equivalent to the solution with $a^2 > 0$ (${\rm AdS}^2$ black hole) these two solutions are globally different due to the behavior of the dilaton field which effectively sets the boundary conditions on the spacetime.

\item ${\rm AdS}_-$: The solution with the value $a^2< 0$ corresponds to a negative mass and although this makes sense in two dimensions (it corresponds to no naked singularities) it will translate in four dimensions into a naked singularity which is unacceptable by cosmic censorship. Hence the solution with $a^2< 0$ is unphysical (from the four-dimensional point of view) and should be discarded.
\end{itemize}

In summary, ${\rm AdS}^2$ black hole (the solution with $a^2> 0$) is characterized by a horizon at $r_H=a/\Lambda$ and a boundary at $r=0$. For the semi-classical process of Hawking radiation the boundary at $r=0$ is not required and therefore one can work in a system of coordinates where the boundary is not accessible. We introduce then the light-cone coordinates

\begin{eqnarray}
\sigma^{\pm}=\tau\pm \sigma.
\end{eqnarray}
Obviously, $\sigma$ is the tortoise coordinate $r_*$ defined as usual by the requirement

\begin{eqnarray}
(\Lambda^2 r^2-a^2)dr_*^2=\frac{dr^2}{\Lambda^2 r^2-a^2}.
\end{eqnarray}
The light-cone coordinates $\sigma^{\pm}$ are given explicitly by (\ref{dilaton1}).

Equivalently, we can work in the light-like coordinates $x^{\pm}$ defined by

\begin{eqnarray}
x^{\pm}=\frac{2}{a\Lambda}\tanh \frac{a\Lambda}{2}\sigma^{\pm}.\label{coc}
\end{eqnarray}
The metric and the dilaton fields in the light-cone and light-like coordinates take the form (conformal gauge)

\begin{eqnarray}
ds^2&=&-\frac{a^2}{\sinh^2\frac{a\Lambda}{2}(\sigma^--\sigma^+)}d\sigma^-d\sigma^+\nonumber\\
&=&-\frac{4}{\Lambda^2}\frac{1}{(x^--x^+)^2}dx^-dx^+.\label{meme}
\end{eqnarray}

\begin{eqnarray}
e^{2(\phi-\phi_0)}&=&\frac{1}{a}\tanh\frac{a\Lambda}{2}(\sigma^--\sigma^+)\nonumber\\
&=&\frac{\Lambda}{2}\frac{x^--x^+}{1-\frac{a^2\Lambda^2}{4}x^-x^+}.
\end{eqnarray}
The ${\rm AdS}^2$ spacetime corresponds thus to setting $a^2=0$ or equivalently $x^{\pm}=\sigma^{\pm}$ in these expressions. In other words, the coordinates $x^{\pm}$ can be thought of as describing ${\rm AdS}^2$ spacetime even for $a^2\ne 0$ since they can be easily extended to the whole of spacetime. We also observe that the boundary of ${\rm AdS}^2$ spacetime is located at $x^-=x^+$ and that we must have $x^-\geq x^+$ (corresponding to $r\geq 0$ in the Schwarzschild coordinates) in order for the dilaton field $\exp(2(\phi-\phi_0))$ to remain positive. Furthermore, it is clear that the coordinates $\sigma^{\pm}$ for $a^2> 0$ cover only the region $-2/a\Lambda< x^{\pm}< +2/a\Lambda$ of the  ${\rm AdS}^2$ spacetime corresponding to the solution $a^2=0$ in the conformal gauge. This region corresponds to the region $r> r_H$ in the Schwarzschild coordinates whereas the boundary at $r=0$ in the Schwarzschild coordinates corresponds now to the line $1-\frac{a^2\Lambda^2}{4}x^-x^+=0$.

Another  interesting system of coordinates consists of  the Poincare coordinates ${t}$ and $z$ defined for the ${\rm AdS}^2$ black hole by the change of coordinates

\begin{eqnarray}
&&{t}=\frac{1}{a\Lambda}e^{a\Lambda \tau}\cosh a\Lambda \sigma\longrightarrow \tau+\frac{1}{a\Lambda}~,~a\longrightarrow 0\nonumber\\
&&z=-\frac{1}{a\Lambda}e^{a\Lambda \tau}\sinh a\Lambda \sigma\longrightarrow -\sigma~,~a\longrightarrow 0.
\end{eqnarray}
The metric in the Poincare patch is given by the usual form

\begin{eqnarray}
ds^2=\frac{1}{\Lambda^2 z^2}(-d{t}^2+dz^2).
 \end{eqnarray}
For $a=0$ (pure ${\rm AdS}^2$ spacetime) the boundary is located at $z=0$  or equivalently $x^--x^+=0$ and the Poincare patch covers $z> 0$ or equivalently $x^--x^+> 0$. This result shows also that the ${\rm AdS}^2$ black hole is indeed locally equivalent to a pure ${\rm AdS}^2$ spacetime. In fact the difference between them is fully encoded in the value of the dilaton field which reflects the boundary conditions imposed on the spacetime and its consequent  topological features.

The relationship between pure ${\rm AdS}^2$ spacetime  ${\rm AdS}_0$ corresponding to the solution $a^2=0$ and the ${\rm AdS}^2$ black hole  ${\rm AdS}_+$ corresponding to the solution $a^2> 0$ is identical to the relationship between the the two-dimensional Minkowski spacetime and the two-dimensional Rindler wedge which are given respectively by the metric tensors

\begin{eqnarray}
ds^2=-dt^2+dz^2.
\end{eqnarray}

\begin{eqnarray}
ds^2=\exp(2\alpha \sigma)(-d\tau^2+d\sigma^2).
\end{eqnarray}
The Rindler wedge is confined to the quadrant $z> |t|$ of Minkowski spacetime  (with $-\infty<\tau,\sigma<+\infty$). The change of coordinates $(t,z)\longrightarrow (\tau,\sigma)$ is given explicitly by

\begin{eqnarray}
t=\frac{1}{\alpha}\exp(\alpha\sigma)\sinh \alpha\tau~,~z=\frac{1}{\alpha}\exp(\alpha\sigma)\cosh \alpha\tau~, ~z>|t|.\label{cha}\nonumber\\
\end{eqnarray}
The asymptotic behavior of the ${\rm AdS}^2$ black hole ${\rm AdS}_+$ is given by the pure ${\rm AdS}^2$ spacetime ${\rm AdS}_0$ which can be fully covered by the light-like coordinates $x^{\pm}$. The change of coordinates (\ref{coc}) which relates the two sets of coordinates $x^{\pm}$ and $\sigma^{\pm}$  (although in our case it does not correspond to any motion of physical observers and therefore is connecting two different manifolds) play exactly the role of the boost which connects the Minkwoski coordinates $(t,x)$ to the Rindler coordinates $(\eta,\sigma)$.



We will use in the following a slightly different parametrization of ${\rm AdS}_+$ and ${\rm AdS}_0$ made possible by the $SL(2,R)$ symmetry of the metric (\ref{meme}) given by the transformations

\begin{eqnarray}
x^{\pm}\longrightarrow \frac{ax^{\pm}+b}{cx^{\pm}+d}~,~ad-bc=1.
\end{eqnarray}
From the first line of (\ref{meme}) the metric on the ${\rm AdS}^2$ black hole ${\rm AdS}_+$ is given by (\ref{coc0}). 
The metric on the ${\rm AdS}^2$ spacetime ${\rm AdS}_0$ is assumed to be of the Poincare form (\ref{coc2}). 
 The change of coordinates $(t,z)\longrightarrow (\tau,\sigma)$ is given explicitly by (\ref{coc3}).

By comparing (\ref{coc1}) with (\ref{cha}) we can see that $a\Lambda$ in our anti-de Sitter black hole plays exactly the role of the acceleration $\alpha$ in Rindler spacetime.
 Equivalently, the parameter $a^2$  (which is proportional to the mass of the black hole) acts effectively as the analogue of the acceleration $\alpha$ with which the Rindler observer is uniformly accelerating in Minkowski spacetime.

These coordinates define region or quadrant I of ${\rm AdS}_0$ (the exterior of our black hole) in which the time-like Killing vector field $\partial_{\tau}$ is future-directed.  
 The region or quadrant II of ${\rm AdS}_0$ in which the time-like Killing vector field (given by $\partial_{-\tau}=-\partial_{\tau}$) is past-directed is defined similarly by equations (\ref{coc4}) and (\ref{coc5}). 


Hence ${\rm AdS}^+$ covers the region of ${\rm AdS}_0$ given by the union of the two quadrants I and II which is specified by the condition $z^+z^-\geq 0$ with the black horizon defined by the condition $z^+z^-=0$.

The equation of motion is the Klein-Gordon equation in the ${\rm AdS}^2$ black hole background ${\rm AdS}_+$ which is locally equivalent to the ${\rm AdS}^2$ spacetime ${\rm AdS}_0$, i.e. the equation of motion is effectively the Klein-Gordon equation in anti-de Sitter spacetime ${\rm AdS}^2$. Furthermore, the inner product  between two solutions $\phi_1$ and $\phi_2$ of the equation of motion is defined in the usual way by ($\Sigma$ is the space-like surface $\tau=0$ and $n^{\mu}$ is the time-like unit vector normal to it)

\begin{eqnarray}
(\phi_1,\phi_2)&=&-i\int_{\Sigma} \big(\phi_1\partial_{\mu}\phi_2^*-\partial_{\mu}\phi_1.\phi_2^*\big) d\Sigma n^{\mu}\nonumber\\
&=&-i\int \big(\phi_1\partial_{\tau}\phi_2^*-\partial_{\tau}\phi_1.\phi_2^*\big) d\sigma.
\end{eqnarray}
A positive-frequency normalized plane wave solution of this equation of motion in region I ($z> 0$) is given by (with $\omega=|k|$)

 \begin{eqnarray}
&&g_k^{(1)}=\frac{1}{\sqrt{4\pi \omega}}\exp(-i\omega \tau+ik\sigma)~,~{\rm I}\nonumber\\
&&g_k^{(1)}=0~,~{\rm II}.\label{pos}
\end{eqnarray}
This is positive-frequency since $\partial_{\tau}g_k^{(1)}=-i\omega g_k^{(1)}$.

A positive-frequency normalized plane wave solution in region II is instead given by

 \begin{eqnarray}
&&g_k^{(2)}=0~,~{\rm I}\nonumber\\
&&g_k^{(2)}=\frac{1}{\sqrt{4\pi \omega}}\exp(i\omega \tau+ik\sigma)~,~{\rm II}.
\end{eqnarray}
Since $\partial_{-\tau}g_k^{(2)}=-i\omega g_k^{(2)}$.

A general solution of the Klein-Gordon equation takes then the form

 \begin{eqnarray}
\phi=\int_k \big(\hat{b}_k^{(1)}g_k^{(1)}+\hat{b}_k^{(2)}g_k^{(2)}+{\rm h.c}\big).
\end{eqnarray}
This should be contrasted with the expansion of the same solution in terms of the anti-de Sitter spacetime modes $f_k\propto \exp(-i(\omega t-kz))$ with $\omega=|k|$ which we will write as

\begin{eqnarray}
\phi=\int_k \big(\hat{a}_k^{}f_k^{}+{\rm h.c}\big).
\end{eqnarray}
The ${\rm AdS}_0$ vacuum $|0_0\rangle$ and the ${\rm AdS}_+$ vacuum $|0_+\rangle$ are defined obviously by

\begin{eqnarray}
\hat{a}_k|0_0\rangle=0.
\end{eqnarray}

\begin{eqnarray}
\hat{b}_k^{(1)}|0_+\rangle=\hat{b}_k^{(2)}|0_+\rangle=0.
\end{eqnarray}
In order to compute the corresponding Bogolubov coefficients we extend the positive-frequency modes $g_k^{(1)}$ and $g_k^{(2)}$ to the entire spacetime ${\rm AdS}_0$ thus replacing the corresponding annihilation operators $\hat{b}_k^{(1)}$ and $\hat{b}_k^{(2)}$ by new annihilation operators $\hat{c}_k^{(1)}$ and $\hat{c}_k^{(2)}$ which annihilate the anti-de Sitter  spacetime vacuum $|0_0\rangle$ \cite{Unruh:1976db}.

Clearly, for $k>0$ we have in region I the following behavior

\begin{eqnarray}
\sqrt{4\pi\omega}g_k^{(1)}&=&\exp(-i\omega\sigma^+)\nonumber\\
&=&(a\Lambda)^{-i\frac{\omega}{a\Lambda}}(z^+)^{-i\frac{\omega}{a\Lambda}}.
\end{eqnarray}
In region II ($z<0$) we should instead consider

\begin{eqnarray}
\sqrt{4\pi\omega}g_{-k}^{(2)*}&=&\exp(-i\omega \sigma^+)\nonumber\\
&=&(-a\Lambda)^{-i\frac{\omega}{a\Lambda}}(z^+)^{-i\frac{\omega}{a\Lambda}}\nonumber\\
&=&e^{\frac{\pi\omega}{a\Lambda}}(a\Lambda)^{-i\frac{\omega}{a\Lambda}}e^{\frac{\pi \omega}{a\Lambda}}(z^+)^{-i\frac{\omega}{a\Lambda}}.
\end{eqnarray}
Thus for all $z$, i.e. along the surface $t=0$, we should consider for $k>0$ the combination

\begin{eqnarray}
\sqrt{4\pi\omega}\big(g_k^{(1)}+e^{-\frac{\pi\omega}{a\Lambda}}g_{-k}^{(2)*}\big)
&=&(a\Lambda)^{-i\frac{\omega}{a\Lambda}}(z^+)^{-i\frac{\omega}{a\Lambda}}.\nonumber\\
\end{eqnarray}
A normalized analytic extension to the entire spacetime of the positive-frequency modes $g_k^{(1)}$ is given by the modes

\begin{eqnarray}
h_k^{(1)}&=&\frac{1}{\sqrt{2\sinh \frac{\pi\omega}{a\Lambda}}}\big(e^{\frac{\pi\omega}{2a\Lambda}} g_k^{(1)}+e^{-\frac{\pi\omega}{2a\Lambda}}g_{-k}^{(2)*}\big).\label{h1}
\end{eqnarray}
Similarly, a normalized analytic extension to the entire spacetime of the positive-frequency modes $g_k^{(2)}$ is given by the modes

\begin{eqnarray}
h_k^{(2)}&=&\frac{1}{\sqrt{2\sinh \frac{\pi\omega}{a\Lambda}}}\big(e^{\frac{\pi\omega}{2a\Lambda}} g_k^{(2)}+e^{-\frac{\pi\omega}{2a\Lambda}}g_{-k}^{(1)*}\big).\label{h2}
\end{eqnarray}
The field operator can then be expanded in these modes as

\begin{eqnarray}
\phi=\int_k \big(\hat{c}_k^{(1)}h_k^{(1)}+\hat{c}_k^{(2)}h_k^{(2)}+{\rm h.c}\big).
\end{eqnarray}
Obviously, the modes $h_k^{(1)}$ and $h_k^{(2)}$ share with $f_k$ the same anti-de Sitter spacetime vacuum $|0_0\rangle $, viz

\begin{eqnarray}
\hat{c}_k^{(1)}|0_0\rangle=\hat{c}_k^{(2)}|0_0\rangle =0.
\end{eqnarray}
The ${\rm AdS}_+$ number operator in region I is defined by

\begin{eqnarray}
\hat{N}_R^{(1)}(k)=\hat{b}_k^{(1)+}\hat{b}_k^{(1)}.
\end{eqnarray}
We can now immediately compute the expectation value of this number operator in region I in the anti-de Sitter vacuum $|0_0\rangle$ to find

\begin{eqnarray}
\langle 0_0|\hat{N}_R^{(1)}(k)|0_0\rangle&=&\langle 0_0|\hat{b}_k^{(1)+}\hat{b}_k^{(1)}|0_0\rangle\nonumber\\
&=&\frac{e^{-\frac{\pi\omega}{a\Lambda}}}{2\sinh\frac{\pi\omega}{2}}\langle 0_0|\hat{c}_{-k}^{(2)}\hat{c}_{-k}^{(2)+}|0_0\rangle\nonumber\\
&=&\frac{1}{e^{\frac{2\pi\omega}{a\Lambda}}-1}\delta(0).
\end{eqnarray}
This is a blackbody Planck spectrum corresponding to the temperature

\begin{eqnarray}
T=\frac{a\Lambda}{2\pi}.
\end{eqnarray}

\section{Note on deformation quantization}
It is well known from deformation quantization that Poisson brackets define the leading correction in the formal expansion of the star product. Explicitly, we should have (recall that upon quantization $i\{f,g\}$ becomes the commutator $[\hat{f},\hat{g}]$)
     \begin{eqnarray}
       f*g(x)&=&f(x)g(x)+\frac{i}{2}\{f,g\}+O(\theta^2)\nonumber\\
       &=&f(x)g(x)+\frac{i}{2}\theta^{ab}\partial_af\partial_bg+O(\theta^2).
     \end{eqnarray}
          On a quantized Poisson manifold  such as the noncommutative ${\rm AdS}^2_{\theta}$ the Poisson structure $\theta$ is generally not a constant tensor and hence the star product is given by   a formal power series in $\theta^{ab}(x)$. The star product which is associative upto the second order in $\theta^{ab}(x)$  is given explicitly by \cite{Kontsevich:1997vbv2}

\begin{eqnarray}
  f*g&=&f.g+\frac{i}{2}\theta^{ab}\partial_af\partial_bg-\frac{1}{8}\theta^{ab}\theta^{cd}\partial_a\partial_cf\partial_b\partial_dg\nonumber\\
  &-&\frac{1}{12}\theta^{ab}\partial_b\theta^{cd}(x)\big(\partial_a\partial_cf\partial_dg+\partial_a\partial_cg\partial_df\big)+O(\theta^3).\label{star1}\nonumber\\
\end{eqnarray}
The noncommutativity tensor $\theta(x)$ satisfies by construction the Jacobi identity
\begin{eqnarray}
  \theta^{ab}\partial_b\theta^{cd}+\theta^{db}\partial_b\theta^{ac}+\theta^{cb}\partial_b\theta^{da}=0.
\end{eqnarray}
The Kontsevich formula for deformation quantization actually only deals with equivalence classes of star products, i.e. a  given star product is really only defined modulo  gauge transformations \cite{Kontsevich:1997vbv2,Cattaneo:1999fm,graciasaz}.

In this context a gauge transformation is a differential linear operator $D~:~A\longrightarrow A$ (where $A$ is the algebra of functions on ${\rm AdS}^2$) which transforms the star product $*$ as follows
\begin{eqnarray}
  && D(f*g)=\bar{f}\bar{*}\bar{g}\equiv \overline{f*g}.\label{definition}
\end{eqnarray}
Where $\bar{*}$ is the gauge transformed star product and functions (elements of the algebra $A$) are transformed as
\begin{eqnarray}
  &&  \bar{f}\equiv D(f)=\sum_nD_n(f).
\end{eqnarray}
The $D_n:A\longrightarrow A$ are differential linear operators of order $O(\theta^n)$. The operator $D$ is invertible (we assume that $D_0=1$) with inverse $E=\sum_nE_n$ (with $E_0=1$) given by \cite{graciasaz}
\begin{eqnarray}
E_n=-\sum_{m=0}^{n-1}E_mD_{n-m}.
\end{eqnarray}
We also write the star products $*$ and $\bar{*}$ as
\begin{eqnarray}
  && f*g=\sum_nB_n(f,g)~,~\bar{f}\bar{*}\bar{g}=\sum_nC_n(\bar{f},\bar{g}).
\end{eqnarray}
The $B_n$ and $C_n$ are bi-differential operators, i.e. $B_n:A\times A\longrightarrow A$ and $C_n:A\times A\longrightarrow A$ of order $O(\theta^n)$. Clearly, the operators $B_n$ are given by (upto the third order $O(\theta^3)$ in the noncommutativity parameter)  
\begin{eqnarray}
  B_0(f,g)&=&fg\nonumber\\
  B_1(f,g)&=&\frac{i}{2}\theta^{ab}\partial_af\partial_b g\nonumber\\
  B_2(f,g)&=&-\frac{1}{8}\theta^{ab}\theta^{cd}\partial_a\partial_cf\partial_b\partial_dg\nonumber\\
  &-&\frac{1}{12}\theta^{ab}\partial_b\theta^{cd}\big(\partial_a\partial_cf\partial_dg+\partial_a\partial_cg\partial_df\big).\label{Be}\nonumber\\
\end{eqnarray}
The definition of the gauge transformation (\ref{definition}) reads then in more detail as (by giving the gauge transformed operators $C_n$ in terms of the $B_k$'s with $k\leq n$)
\begin{eqnarray}
C_n(\bar{f},\bar{g})=\sum_{m+k+l+j=n} D_m(B_k(E_l\bar{f},E_j\bar{g})).
\end{eqnarray}
We compute explicitly the first three terms:
\begin{eqnarray}
  C_0(\bar{f},\bar{g})=\bar{f}\bar{g}.
\end{eqnarray}
\begin{eqnarray}
  C_1(\bar{f},\bar{g})=B_1(\bar{f},\bar{g})+D_1(\bar{f}\bar{g})-(D_1\bar{f})\bar{g}-\bar{f}(D_1\bar{g}).\label{C1e}\nonumber\\
\end{eqnarray}
\begin{eqnarray}
  C_2(\bar{f},\bar{g})&=&B_2(\bar{f},\bar{g})+D_2(\bar{f}\bar{g}) -(D_2\bar{f})\bar{g}-\bar{f}(D_2\bar{g})\nonumber\\
  &-&D_1((D_1\bar{f})\bar{g})-D_1(\bar{f}(D_1\bar{g}))+(D_1\bar{f})(D_1\bar{g})\nonumber\\
  &+&(D_1^2\bar{f})\bar{g}+\bar{f}(D_1^2\bar{g})+D_1(B_1(\bar{f},\bar{g}))\nonumber\\
  &-&B_1(D_1\bar{f},\bar{g})-B_1(\bar{f},D_1\bar{g}).\label{C2e}
\end{eqnarray}
Thus, upto the third order $O(\theta^3)$ in the noncommutativity parameter the gauge transformation (\ref{definition}) reads 
\begin{eqnarray}
D(f*g)&=&{f}{*}{g}+D_1(fg)+D_1(B_1(f,g))+D_2(fg)\nonumber\\
&+&O(\theta^3).
\end{eqnarray}
All star products in the Kontsevich theory of deformation quantization are determined in terms of Poisson structures while gauge transformations correspond typically to coordinate transformations $x\longrightarrow \bar{x}$. But under a coordinate transformation $x\longrightarrow \bar{x}$ we must obviously have
\begin{eqnarray}
&&\theta^{ab}\longrightarrow   \bar{\theta}^{ab}=\theta^{cd}(x)\frac{\partial \bar{x}^a}{\partial x^c}\frac{\partial \bar{x}^b}{\partial x^d}\Rightarrow  \nonumber\\
&&B_1(f,g)=\frac{i}{2}\bar{\theta}^{ab}\bar{\partial}_af\bar{\partial}_b g.
\end{eqnarray}
Thus,  the Poisson structure in the transformed coordinates $\bar{x}$ is precisely given by the noncommutativity parameter $\bar{\theta}^{ab}$ and therefore the corresponding star product is given by $\bar{*}$, i.e. 
\begin{eqnarray}
C_1({f},{g})=  B_1(f,g)=\frac{i}{2}\bar{\theta}^{ab}\bar{\partial}_af\bar{\partial}_b g.
\end{eqnarray}
By comparing this result with  equation (\ref{C1e}) we conclude that the gauge transformation $D_1$ (if it corresponds to the coordinate transformation $x\longrightarrow \bar{x}$) must satisfy the Leibniz rule, viz
\begin{eqnarray}
  D_1(\bar{f}\bar{g})=(D_1\bar{f})\bar{g}+\bar{f}(D_1\bar{g}).\label{lei}
\end{eqnarray}
By using this result in (\ref{C2e}) we obtain
\begin{eqnarray}
  C_2(\bar{f},\bar{g})&=&B_2(\bar{f},\bar{g})+D_2(\bar{f}\bar{g}) -(D_2\bar{f})\bar{g}-\bar{f}(D_2\bar{g})\nonumber\\
  &-&(D_1\bar{f})(D_1\bar{g})+D_1(B_1(\bar{f},\bar{g}))\nonumber\\
  &-&B_1(D_1\bar{f},\bar{g})-B_1(\bar{f},D_1\bar{g}).
\end{eqnarray}
A simple solution is given by $D_1=0$, i.e. the gauge transformation starts to become non-trivial only at the second order $O(\theta^2)$ in the noncommutativity parameter.

\end{document}